\documentclass[12pt,a4paper]{article}
\usepackage[utf8]{inputenc}
\usepackage{fullpage}
\usepackage{hyperref}
\usepackage{amssymb}
\usepackage{amsthm}
\usepackage{graphicx}
\usepackage[outdir=./]{epstopdf}
\usepackage{color,soul}
\usepackage{xcolor}
\usepackage{soul}

\usepackage{enumitem}
\usepackage{empheq}
\usepackage{cite}
\usepackage{slashed}
\usepackage{amsmath}
\usepackage{braket}
\usepackage{simpler-wick}
\usepackage{simplewick}

\numberwithin{equation}{section}
\usepackage{pgfplots}
\ifdim\overfullrule>0pt 
\usepackage{environ}

\NewEnviron{tikzpicture}{%
	\begin{pgfpicture}
		\pgfpathrectanglecorners{\pgfpointorigin}{\pgfpoint{3cm}{3cm}}%
		\pgfusepath{stroke}\end{pgfpicture}%
}
\fi

\renewcommand{\Re}{\operatorname{Re}}

\usepackage{cleveref}
\crefname{section}{Section}{Sections}
\crefname{appendix}{Appendix}{Appendices}

\begin{document}

\titlepage

\begin{flushright}
	MS-TP-21-37
\end{flushright}

\vspace*{1.2cm}

\begin{center}
	{\Large \bf Classical and quantum gravitational scattering with 
		\\[0.5cm]
	Generalized Wilson Lines}
	
	\vspace*{1.5cm} \textsc {D. Bonocore, A. Kulesza, J. Pirsch} \\
	
	\vspace*{1cm}

Institut f\"{u}r Theoretische Physik, Westf\"{a}lische
	Wilhelms-Universit\"{a}t M\"{u}nster, Wilhelm-Klemm-Stra\ss e 9,
	D-48149 M\"{u}nster, Germany\\

\end{center}

\vspace*{7mm}

\begin{abstract}
	\noindent 
The all-order structure of scattering amplitudes
 is greatly simplified by the use of  
Wilson line operators, describing eikonal emissions from straight lines 
extending to infinity. 
A generalization at
subleading powers in the eikonal expansion, known as Generalized Wilson Line 
(GWL), has been proposed some time ago, and has been applied both in QCD 
phenomenology and in the high energy limits of
gravitational amplitudes.
In this paper we revisit the construction of the scalar gravitational GWL 
starting from first 
principles in the worldline formalism. We identify the correct 
Hamiltonian that leads to a simple correspondence between the soft expansion 
and the weak field expansion. This allows us to isolate the terms in the GWL 
that are relevant in
the classical limit. 
In doing so we devote special care to the regularization of UV divergences that 
were not 
discussed in an earlier derivation. 
We also clarify the relation with  
a parallel body of work that recently investigated the classical limit of 
scattering 
amplitudes in gravity in the worldline formalism.
	
\end{abstract}


\newpage

\section{Introduction}
\label{sec:intro}

The rise of gravitational wave astronomy, and the subsequent increasing demand 
for highly precise theoretical predictions, have attracted a great deal of 
interest in the scattering amplitude community over the last years. There is 
growing evidence that
the rich toolbox developed to investigate the scattering of elementary 
particles provides a useful framework to investigate binary inspirals and 
mergers of 
compact astrophysical objects such as black-holes and neutron stars. 
The progress made on this front in the recent years is remarkable, and relies 
on the profound observation that it is possible to gain new insights into 
classical 
scattering by studying the (apperently more challenging) quantum counterpart 
\cite{Bern:2020buy, Kosower:2018adc, Maybee:2019jus, Cristofoli:2021vyo, 
	Cheung:2018wkq, Goldberger:2004jt, Dlapa:2021npj, Kalin:2020fhe, 
	Damour:2019lcq, Damour:2017zjx, tHooft:1987vrq, Amati:1990xe, Amati:1987uf, 
	Ciafaloni:2018uwe, 
	Ciafaloni:2015vsa, 
	DiVecchia:2021bdo, DiVecchia:2021ndb, DiVecchia:2020ymx, DiVecchia:2019kta, 
	DiVecchia:2019myk, Heissenberg:2021tzo, Damgaard:2021ipf, 
	Bjerrum-Bohr:2018xdl, 
	Bjerrum-Bohr:2021din, KoemansCollado:2018hss, KoemansCollado:2019ggb, 
	Giddings:2010pp, 
	AccettulliHuber:2020oou, Bern:2019nnu, Bern:2019crd, Bern:2020gjj, 
	Parra-Martinez:2020dzs, 
	Kabat:1992tb, 
	Akhoury:2013yua, Bjerrum-Bohr:2016hpa, Brandhuber:2021eyq, 
	Brandhuber:2021kpo}.

Among the various strategies that have been pursued in this program,
worldline methods have received a renewed attention in the recent years  
\cite{Levi:2015msa, Goldberger:2016iau, Chester:2017vcz, 
Shen:2018ebu, 
Plefka:2018dpa,  
Almeida:2020mrg, 
Bastianelli:2021rbt, Corradini:2021jgd, Riva:2021vnj}.
In particular, Mogull, 
Plefka and Steinhoff \cite{Mogull:2020sak, Jakobsen:2021smu, Jakobsen:2021lvp, 
Jakobsen:2021zvh} recently proposed a method, named Worldline Quantum Field 
Theory (WQFT), 
 to compute Post-Minkowskian corrections to classical observables. Building on 
 the well-established string-inspired formalism  
that represents scattering 
amplitudes in terms of first-quantized path integrals \cite{Schubert:2001he}, 
this approach follows 
from the  
representation of dressed asymptotic states in terms of path integrals over 
the trajectory of the colliding massive objects.

A similar description was proposed some time 
ago 
in gauge theories
 by 
Laenen, 
Stavenga and White \cite{Laenen:2008gt} and later extended to gravity by White 
\cite{White:2011yy}. The asymptotic dressed propagator constructed in this 
approach has been dubbed a Generalized 
Wilson line (GWL), since it describes asymptotic states 
of a (quantum) scattering amplitude dressed by radiation at higher orders in 
the 
soft expansion, thus generalizing the representation of scattering 
amplitudes 
with Wilson lines. 
It has found applications both in QCD phenomenology 
\cite{Bonocore:2016awd, Bahjat-Abbas:2019fqa} and in the Regge limit of 
gravitational scattering 
\cite{Luna:2016idw}.

More specifically, the scalar gravitational GWL 
for a straight semi-infinite path along the direction $p^{\mu}$ 
is defined at the next-to-soft level as 
\cite{White:2011yy}\footnote{Note 
that 
	the GWL 
	in \cref{gwl} differs from the one derived in \cite{White:2011yy} 
	because of different conventions. In particular, in this work we define the 
	weak field expansion as in 
	\cref{weak}, which is more standard in contemporary literature.}
\begin{align}
\widetilde W_p(0,\infty)&=\exp\Bigg\{
\frac{i\kappa}{2}\int_0^{\infty} dt\,\left[- p_{\mu}p_{\nu}
+ip_{\nu}\partial_{\mu}
-\frac{i}{2}\eta_{\mu\nu}p^{\alpha}\partial_{\alpha}
+\frac{i}{2}tp_{\mu}p_{\nu}\partial^2\right]h^{\mu\nu}(pt)
 \notag \\
&\qquad + \frac{i\kappa^2}{2}\int_0^{\infty} dt \int_0^{\infty} ds\,
\Bigg[
\frac{p^{\mu}p^{\nu}p^{\rho}p^{\sigma}}{4}\min(t,s)\,\partial_{\alpha}
h_{\mu\nu}(pt)
\partial^{\alpha}
h_{\rho\sigma}(ps)
\notag\\&\qquad
+p^{\mu}p^{\nu}p^{\rho}\,\theta(t-s)\,
h_{\rho\sigma}(ps)
\partial_{\sigma}
h_{\mu\nu}(pt)
\,+\,p^{\nu}p^{\sigma}\,\delta(t-s)\,
h^{\mu}_{\;\,\sigma}(ps)
h_{\mu\nu}(pt)
\Bigg]\Bigg\}~.
\label{gwl}
\end{align} 
Here we defined the graviton $h_{\mu\nu}$ via the
weak field expansion
\begin{align}
g_{\mu\nu}=\eta_{\mu\nu}+\kappa h_{\mu\nu}~,
\label{weak}
\end{align}
where $\kappa^2=32\pi G$ and $G$ is the Newton constant.
The first term in the first line of \cref{gwl} corresponds to a standard Wilson 
line on a straight semi-infinite path, 
while all other terms correspond to next-to-soft (or next-to-eikonal) 
interactions. Note in particular the presence of correlations between two 
graviton fields at different times at this order.

The gravitational GWL has been first  
derived by White in \cite{White:2011yy} by writing a Schwinger representation 
for the 
dressed propagator, obtained from the quadratic part of the corresponding 
scalar 
field theory Lagrangian, after the weak field expansion
 has been performed. 
 Although this procedure leads to the correct result, it is the 
 purpose of this 
 paper to show how to derive the GWL in an elegant way from first 
 principles in the worldline formalism (i.e. 
 from the constrained quantization of the relativistic particle action in 
 a generic curved background). By following this route we
 clarify a few issues that remain ambiguous in the 
other derivation. 

Most notably, when one upgrades the background metric from a flat 
$\eta_{\mu\nu}$ to a generic 
curved spacetime $g_{\mu\nu}$, the worldline action becomes a 
superrenormalizable non-linear sigma model of the type $g_{\mu\nu}\dot 
x^{\mu}\dot x^{\nu}$, which contains ultraviolet (UV) divergences
whose renormalization requires the use of ghosts fields.  
Although this 
feature (not discussed in 
\cite{White:2011yy}) has been extensively investigated in the literature 
 \cite{DeBoer:1995hv, Peeters:1993vu, 
Schubert:2001he, Bastianelli:2002fv, 
Bastianelli:2002qw, Bastianelli:2003bg, Corradini:2021jgd},
it is not immediately clear 
how the 
exponentiated next-to-soft terms in 
\cref{gwl} are affected by the regularization scheme. 
In fact,  
the Hamiltonian $H(x,p)$ needed to construct the worldline path integral is 
uniquely 
defined 
only within a regularization scheme after an ordering prescription for the 
operators $\hat x$ and $\hat p$ has been chosen. 
In the worldline literature $H(x,p)$ is typically defined in
Weyl-ordering, i.e. by symmetrizing all $\hat x$ and $\hat p$ operators. The 
GWL on 
the other hand represents soft radiation emitted from an asymptotic propagator 
of 
well-defined final momentum 
emerging from a localized hard interaction, and 
therefore is more 
naturally derived in $px$-ordering\footnote{Note that 
	this prescription has been unconventionally dubbed Weyl-ordering in 
	\cite{Laenen:2008gt, White:2011yy, Bonocore:2020xuj}.}, 
i.e. with all $\hat p$ operators on the left of $\hat x$ operators 
\cite{Laenen:2008gt, 
White:2011yy, Bonocore:2020xuj}. 
In fact, another motivation  
for this work is to show in detail that a quantization procedure in curved 
spacetime  
with these less common conventions leads indeed to the 
correct 
result for the GWL, at the price of setting up a quantization with 
a non-hermitian conjugate momentum. The resulting Hamiltonian is then much 
simpler than 
the one derived in \cite{White:2011yy} and exhibits a simple connection 
between the weak field expansion and the soft expansion.

Another point that one would like to address is whether the GWL provides a way 
to isolate the classical contribution and discard the quantum part.
This is an aspect not immediately evident 
in the original calculation of \cite{White:2011yy} and \cite{Luna:2016idw}, 
where one has to first carry out the calculation at the amplitude level to 
discover that some diagrams 
are subleading in the Regge limit and therefore contribute only to the Regge 
trajectory and not to the classical eikonal phase. 
We will see that a Hamiltonian that is derived from first principles will 
clarify 
this issue by isolating purely quantum terms in \cref{gwl}.

A third motivation for the present study arises from a comparison with the 
aforementioned 
Worldline Quantum Field Theory formalism \cite{Mogull:2020sak}. 
Indeed, the two approaches follow a similar strategy by 
representing power suppressed 
graviton emissions as path integrals along the worldlines of the massive 
particles. However, naively comparing the Hamiltonian of \cite{White:2011yy}
with the Hamiltonian of \cite{Mogull:2020sak} one is tempted to conclude that 
the two formalisms describe quite different physics. On the other hand, the 
calculation of the eikonal phase 
leads to the same result in both approaches, and includes corrections of order 
$G/z$, where $z$ is 
the impact parameter 
\cite{Luna:2016idw}. Once again, the puzzle 
is solved by deriving the GWL from first principles. In fact, we demonstrate  
 that the GWL and the WQFT 
formalism are equivalent in the classical limit.

Finally, when one attempts to define the 
GWL for 
spinning emitters, a derivation from the constrained quantization of the 
worldline model with Grassmann variables
 becomes mandatory to prove
the exponentiation of numerator contributions. In fact, as already discussed in 
the gauge theory case
\cite{Bonocore:2020xuj}, the supersymmetry of the worldline model guarantees 
that the 
background field in the numerator of the dressed propagator does not contribute 
in the asymptotic limit. Therefore, identifying the correct scalar Hamiltonian
in curved spacetime that yields the desired result is a fundamental step that 
paves the way for the 
spinning case.

The structure of the paper is as follows. We begin in \cref{sec:fact} by 
reviewing the factorization of next-to-soft emissions, to stress that the GWL 
is a 
generalization to all order of that procedure. In \cref{sec:review} we 
revisit the derivation of the GWL in White's approach \cite{White:2011yy}, i.e. 
with  the  
Schwinger representation of the dressed propagator after the weak field 
expansion has been performed. Then, in \cref{sec:worldline} we discuss the 
derivation
of the GWL from a worldline model in a generic curved background. In doing so, 
we identify the 
correct Hamiltonian in $px$-ordering leading to the correct next-to-soft 
emissions
of \cref{sec:fact}. Finally, 
in \cref{sec:ampl} we discuss applications at the amplitude level in 
the classical limit and we compare the GWL and the WQFT approaches.
Technical details underlying the calculations of \cref{sec:review} and 
\cref{sec:worldline}
are presented in separate appendices.

\section{Factorization of next-to-soft emissions}
\label{sec:fact}

We consider 
a complex scalar field minimally coupled to gravity in $d$-dimensions via
\begin{align}
S=\int d^dx\, \sqrt{-g}\left(
g^{\mu\nu}\partial_{\mu}\phi^* \partial_{\nu}\phi 
-m^2|\phi|^2\right)
~,
\label{Spm}
\end{align}
where 
$g$ is 
the determinant of the metric $g_{\mu\nu}$. Note that throughout this paper we 
use a mostly minus metric.
We define the weak 
field expansion via \cref{weak}.
 In the following we will need to include terms up to order 
$\kappa^2$ for a consistent next-to-soft expansion. Thus, we need also 
\begin{align}
g^{\mu\nu}&=\eta^{\mu\nu}-\kappa h^{\mu\nu}+\kappa h^{\mu\rho}h^{\nu}_{\,\,\rho}
+{\cal O}(\kappa^3)~,
\label{weakinv} \\
\sqrt{-g}&=1+\frac{\kappa}{2}h+\frac{\kappa^2}{4}\left(\frac{h^2}{2}-h^2_{\mu\nu}\right)+{\cal
 O}(\kappa^3)~,
\label{trace}
\end{align}
where $h=\eta^{\mu\nu}h_{\mu\nu}$. Then, the Feynman rules
for single and double graviton emissions can be easily extracted from 
\cref{Spm} and read
\begin{align}
&V_{\mu\nu}=i\frac{\kappa}{2}\left(p_\mu p'_\nu+p'_\mu 
p_\nu-\eta_{\mu\nu}(pp'+m^2)\right)~,\label{V1grav}\\
&V_{\mu\nu\rho\sigma}=i\frac{\kappa^2}{4}\Big[(pp'+m^2)(-\eta_{\mu\nu}\eta_{\rho\sigma}
+\eta_{\mu\rho}\eta_{\nu\sigma}+\eta_{\nu\rho}\eta_{\mu\sigma}) 
+\eta_{\mu\nu}(p'_\rho p_\sigma+p_\rho 
p'_\sigma)+\eta_{\rho\sigma}(p_\mu p'_\nu+p'_\mu p_\nu) \notag\\
& -\left(\eta_{\mu\rho}(p'_\sigma p_\nu+p'_\nu 
p_\sigma)+\eta_{\nu\rho}(p'_\sigma p_\mu+p'_\mu p_\sigma)\right)
-\left(\eta_{\mu\sigma}(p'_\rho p_\nu+p'_\nu 
p_\rho)+\eta_{\nu\sigma}(p'_\rho p_\mu+p'_\mu 
p_\rho)\right)\Big]\label{V2grav}~,
\end{align}
where we assumed all momenta to be incoming.

\begin{figure}
	\centering
	\includegraphics[width=50mm]{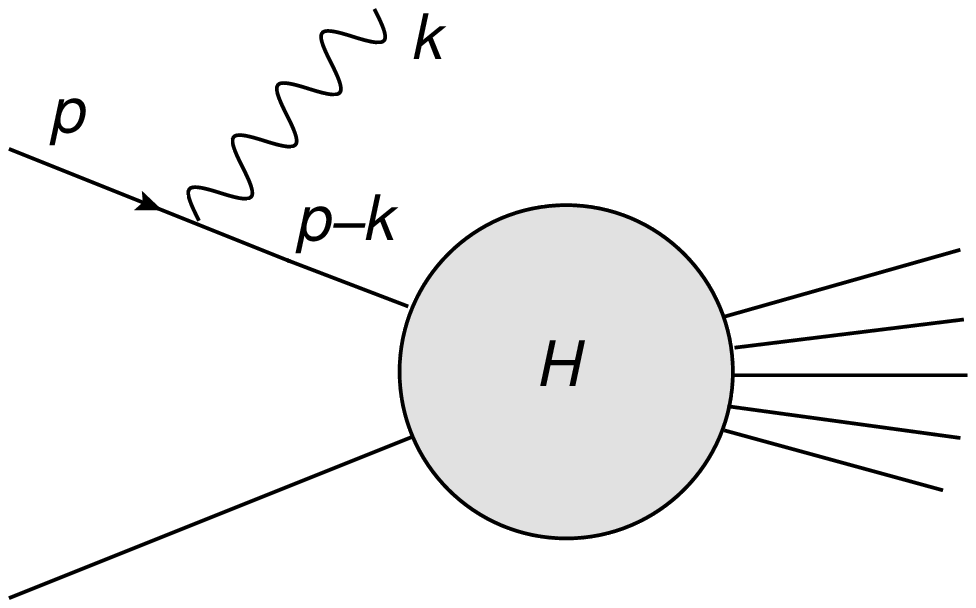}
	\qquad 
	\includegraphics[width=50mm]{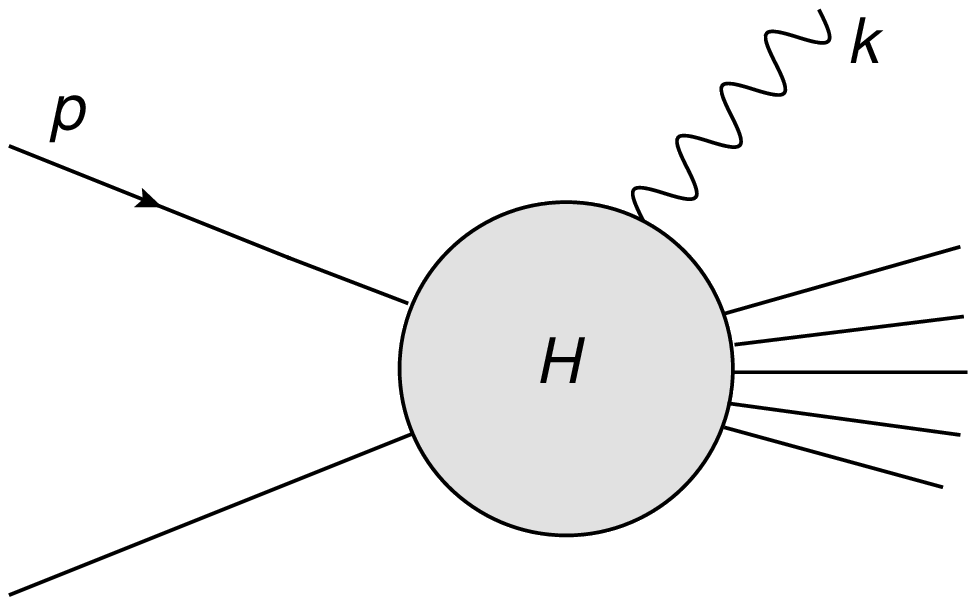}
	\caption{Diagrammatic representation of an external emission (left) and 
		internal emission (right) in a scattering amplitude. The blob $H$ 
		represents a generic subdiagram sensitive to the unspecified hard 
		dynamics.}
	\label{fig:internal}
\end{figure}

The factorization of two next-to-soft graviton emissions proceeds differently 
for external and internal emissions, as originally presented by Low in the 
gauge theory case \cite{Low:1958sn, White:2014qia, Gervais:2017zky, 
Beneke:2021umj}. The difference between the 
two cases is 
better understood by looking at the diagrams in \cref{fig:internal} for a 
single next-to-soft boson.  

More specifically, for one 
 external emission, one considers a single line of momentum $p_i$ and simply 
expands the diagram at next-to-leading power in the soft graviton momentum $k$. 
This amounts to either expanding the vertices in \cref{V1grav},  
\cref{V2grav} and the scalar propagator of momentum $p_i-k$, or to expanding 
the 
hard function as 
\begin{align}
H(p_1,...,p_i-k,...,p_n)=
H(p_1,...,p_i,...,p_n)-k_{\mu}\frac{\partial}{\partial 
	p_{\mu}}H(p_1,...,p_i,...,p_n)~.
\label{low}
	\end{align}
For an internal emission, on the other hand, one cannot naively compute the 
diagram in the right of \cref{fig:internal} due to the 
ignorance about the coupling of the soft boson with the hard subdiagram. 
Instead one exploits the gauge invariance of the full amplitude 
to express the result in terms of the external emission diagrams. For on-shell 
gravitons, the outcome 
is that 
the internal emission diagram and the derivative term in \cref{low} combine 
into a term consisting of the 
orbital angular momentum $L^{\mu\nu}=p^{\mu}\frac{\partial}{\partial p^{\nu}}
-p^{\nu}\frac{\partial}{\partial p^{\mu}}$ acting on the non-radiative 
amplitude. More generally, one can observe that the internal contribution is 
equal to an eikonal emission dressing the non-radiative amplitude with shifted 
kinematics \cite{Akhoury:2013yua, Luna:2016idw}.

\begin{figure}
	\centering
	\includegraphics[width=130mm]{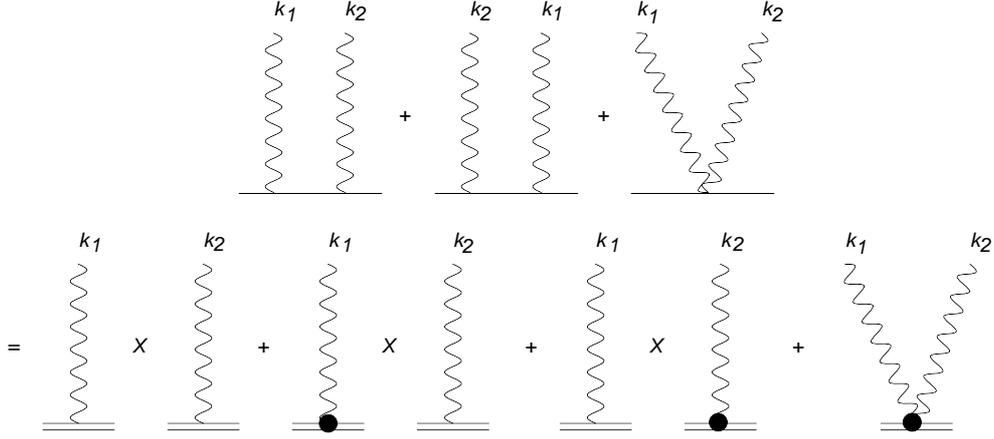}
	\caption{Factorization of two graviton emissions of momenta $k_1$ and $k_2$ 
		at order $\kappa^2$. Diagrams on the l.h.s. are constructed with the 
		full vertices of \cref{V1grav} and \cref{V2grav} and then expanded at 
		next-to soft level. The diagrams on the r.h.s. instead contain
		the combined vertex-propagator Feynman rules of \cref{Evertex}, 
		\cref{NEvertex} and \cref{NEvertices}. Next-to-eikonal emissions are 
		denoted with a blob.}
	\label{NEid}
\end{figure}

The contribution from the external emissions without derivatives acting on $H$ 
can be 
elaborated further. 
For example, for two emissions one 
sums over all possible graviton insertions on the scalar line, as shown in 
\cref{NEid}. While for emissions of order $\kappa$ this 
presents no difficulty, for contributions of order $\kappa^2$ the algebra is 
quite tedious and not 
shown here. Still, 
the result can be written in a  relatively 
 compact form in terms of three vertices:
\begin{align}
V^{\text{E}}_{\mu\nu}(k)&=-\frac{\kappa}{2}\frac{p_\mu 
p_\nu}{pk}~,
 \label{Evertex}\\
 V^{\text{NE}}_{\mu\nu}(k)&=-\frac{\kappa}{4}\frac{k^2 p_\mu 
 	p_\nu}{(pk)^2}+\frac{\kappa}{4}\frac{1}{pk}(k_\mu p_\nu+k_\nu 
 p_\mu-\eta_{\mu\nu}pk)~,
 \label{NEvertex}\\
V^{\text{NE}}_{\mu\nu\rho\sigma}(k,l)&=\frac{\kappa^2}{4}\frac{1}{p(k+l)}\Bigg[\frac{kl}{pk~pl}p_\mu
 p_\nu 
p_\rho p_\sigma-\frac{p_\rho p_\sigma}{pl}(p_\mu l_\nu+p_\nu 
l_\mu)-\frac{p_\mu p_\nu}{pk}(p_\rho k_\sigma+p_\sigma k_\rho)\notag \\
&\qquad +\eta_{\mu\rho}p_\sigma p_\nu+\eta_{\nu\rho}p_\sigma 
p_\mu+\eta_{\mu\sigma}p_\rho p_\nu+\eta_{\nu\sigma}p_\rho p_\mu\Bigg]~,
\label{NEvertices}
\end{align}
where $k$ and $l$ denote the soft momenta. 
The first vertex represents the well-known single-graviton eikonal (E) Feynman 
rule, 
already present in Weinberg's theorem \cite{Weinberg:1965nx}, while the 
next-to-eikonal (NE) correction is given by $V^{\text{NE}}_{\mu\nu}$
\footnote{Note that the contribution from 
$V^{\text{NE}}_{\mu\nu}$ vanishes for on-shell gravitons in 
de Donder gauge (i.e. when  
$\frac{1}{2}\partial^{\mu}h-\partial_{\nu}h^{\mu\nu}=0$), in agreement with the 
next-to-soft theorems of \cite{Cachazo:2014fwa}. However, applications in 
high-energy scattering require off-shell gravitons. In that case, the second 
term in \cref{NEvertex} brings a non-vanishing correction to the Regge 
trajectory \cite{Luna:2016idw}.  }. 
At subleading order we 
should consider also a two-graviton seagull-like vertex, given by
$V^{\text{NE}}_{\mu\nu\rho\sigma}$.

The question is whether this factorization persists at every order in $\kappa$, 
i.e. whether
the sum of all possible attachments of $N$ gravitons emissions at next-to-soft 
level can be written 
exclusively in terms of the vertices of \cref{Evertex}, \cref{NEvertex} and 
\cref{NEvertices}. 
As we are going to discuss, the GWL is meant to implement this idea by 
achieving an exponentiation along 
the hard particle worldline.  

\section{GWL from four-dimensional field theory}
\label{sec:review}
The derivation of the GWL from the field 
theory Lagrangian has been already discussed in detail in 
\cite{White:2011yy} after defining the graviton field via
\begin{align}
\sqrt{g}g_{\mu\nu}=\eta_{\mu\nu}+\kappa h_{\mu\nu}~.
\label{weakg}
\end{align}
In order to highlight the main 
points 
that will be addressed in the worldline approach of 
the next sections, here
we revisit the approach of \cite{White:2011yy} to derive the GWL. In 
particular, we use the 
weak field expansion of 
\cref{weak}, which is more common in the literature, instead of \cref{weakg}.
In fact, it is worth noting that, unlike in gauge theories,  
 gravitational GWLs are sensitive to how the 
graviton field is defined via the 
weak field expansion. Also, we use a mostly minus 
metric throughout.

We begin with the  
Schwinger representation of the dressed propagator. We first take \cref{Spm} 
and perform the weak field expansion with \cref{weak}, \cref{weakinv} and 
\cref{trace}.
 Then, after 
 integrating by parts and neglecting the surface term, 
 \cref{Spm} becomes 
 \begin{align}
 S&=\notag \int d^dx\,\phi^*(x)\Bigg( 
 \Box+m^2 +\kappa\left[-\partial^\mu\partial^\nu 
 h_{\mu\nu}+\partial^\nu(\partial^\mu 
 h_{\mu\nu})+(\Box+m^2)\frac{h}{2}-\frac{1}{2}\partial_\mu(\partial^\mu 
 h)\right]\\
 &\quad 
 +\kappa^2\Bigg[(\Box\!+\!m^2)\left(\frac{h^2}{8}\!-\!\frac{h_{\mu\nu}^2}{4}\right)\!
 -\partial^\mu\left(\!\partial_\mu\!\left(\frac{h^2}{8}\!-\!
 \frac{h_{\mu\nu}^2}{4}\right)\!\right)-\!\partial^\mu
 \partial^\nu\frac{h}{2}h_{\mu\nu}\!\notag \\
 &\quad +\!\partial^\nu\frac{h}{2}(\partial^\mu
  h_{\mu\nu})\!+\partial^\nu\!(\partial^\mu\frac{h}{2}) 
 h_{\mu\nu}+\partial_\mu\partial_\nu h^\mu_\rho 
 h^{\rho\nu}-\partial_\mu(\partial_\nu h^\mu_\rho 
 h^{\rho\nu})\Bigg]\Bigg)\phi(x)~.
 \label{act-exp}
 \end{align}
By writing \cref{act-exp} 
in the form 
$\int d^4x\,
\phi^* (2 H) \phi$, we can define $2 H$ as the inverse scalar propagator 
dressed 
with  
gravitational radiation. The Schwinger representation of this propagator 
consists of 
interpreting $H$ as a Hamiltonian governing the 
evolution 
in proper time $T$ of a relativistic scalar particle. 
 This is achieved by first
replacing all derivatives in $H$ acting on 
a wave function via  $i\partial_{\mu}\to \hat p_{\mu}$.
Hence, in $px$-ordering (i.e. with all 
momentum operators on the left of the position operator) the Hamiltonian reads
\begin{align}
\hat H=&-\frac{1}{2}\Bigg( p^2-m^2+\kappa\left[-p^\mu 
 p^\nu h_{\mu\nu}
 +ip^\nu(\partial^\mu h_{\mu\nu})
 +\frac{1}{2}(p^2-m^2)h-\frac{i}{2}p_\mu(\partial^\mu h)\right] \notag \\
&+\kappa^2\!\Bigg[(p^2-m^2)\left(\frac{h^2}{8}-\frac{h_{\mu\nu}^2}{4}\right)
\!-ip^\mu\partial_\mu\!\left(\frac{h^2}{8}\!-\!\frac{h_{\mu\nu}^2}{4}\right)\!
-\frac{1}{2}p^\mu
p^\nu h_{\mu\nu} h\! \notag \\
&+\!\frac{i}{2}p^\nu h(\partial^\mu 
h_{\mu\nu})\!
+\!\frac{i}{2}p^\nu\left(\partial^\mu h\right) h_{\mu\nu}+p_\mu p_\nu 
h^\mu_\rho h^{\rho\nu}-ip_\mu (\partial_\nu h^\mu_\rho 
h^{\rho\nu})\Bigg]\Bigg) + {\cal O}(\kappa^3)~.
\label{eq:Hamiltonian}
\end{align}
The dressed propagator 
in position representation then becomes 
\begin{align}
\langle{x_f}|(2i(H-i\epsilon))^{-1}|{x_i}\rangle
=\frac{1}{2}\int\limits_{0}
^{\infty}dT~\langle{x_f}|e^{-i(H-i\epsilon)T}|{x_i}\rangle=
\int\limits_{0}^{\infty}dT 
\int\limits\displaylimits_{x(0)=x_i}^{x(T)=x_f}\mathcal{D}x\, \mathcal{D}p\, 
e^{-i\int\limits_{0}^{T}dt(p\dot{x}+H(x,p)-i\epsilon)}~,
\label{schwinger}
\end{align}
where we inserted the standard Feynman $i \epsilon$ prescription, and in the 
second step we used a path integral representation for the matrix elements.

In the following, we will be interested in solving \cref{schwinger} order by 
order in the soft expansion in the asymptotic limit. To do so, we need to  
Fourier transform \cref{schwinger}
\emph{before} carrying out the path integral over $x$. This can be done by  
considering 
the transition element 
 from an initial state of 
position $x_i$ to a final state of momentum $p_f$.
The asymptotic propagator is then defined as the dressed propagator truncated 
of the external free propagator of momentum $p_f$. 
After performing the Gaussian momentum integration in \cref{schwinger} 
and expanding around the 
classical solutions $x=x_i+p_ft+\tilde x$,
the asymptotic propagator reads
\begin{equation}
-i(p_f^2-m^2+i\epsilon) 
\,\bra{p_f}(2i(H-i\epsilon))^{-1}\ket{x_i}
=-i(p_f^2-m^2+i\epsilon) 
\int_0^{\infty} dT\,
e^{ip_fx_i-\frac{i}{2}(p_f^2-m^2+i\epsilon)T}
\, f(T)~.
\label{ft}
\end{equation}
Here we defined  
\begin{align}
f(T)\equiv \int\limits_{\tilde 
	x(0)=0}\mathcal{D}\tilde x 
\exp\left(i\int\limits_{0}^{T}dt\, L[\tilde x(t)]\right)~,
\end{align}
where
\begin{align}
L[x(t)]&\nonumber=-\frac{\lambda\dot{x}^2}{2}+\frac{\kappa}{2}\Bigg[-\lambda 
h_{\mu\nu}p^\mu p^\nu+i\left(\partial^\mu 
h_{\mu\nu}p^\nu-\frac{1}{2}\partial_\mu 
hp^\mu\right)+\\&\nonumber+\lambda\dot{x}^\mu\left(-2h_{\mu\nu}p^\nu+hp_\mu\right)
+i\dot{x}^\mu\left(\partial^\nu
 h_{\mu\nu}-\frac{1}{2}\partial_\mu 
h\right)+\lambda\dot{x}^\mu\dot{x}^\nu\left(\frac{h}{2}\eta_{\mu\nu}-h_{\mu\nu}\right)\Bigg]\\
&\nonumber+\frac{\kappa^2}{4}\Bigg[\lambda \left(hh_{\mu\nu}p^\mu 
p^\nu-\frac{h^2}{2}m^2\right)+i\left(h_{\rho\sigma}\partial^\mu 
h_{\rho\sigma}p_\mu-2p^\nu (\partial^\rho h_{\mu\nu}) 
h_{\mu\rho}\right)+\\
&\nonumber~~~~~~~~+\frac{1}{\lambda}\left(\frac{1}{8}\partial_\mu h\partial^\mu 
h+\frac{1}{2}\partial^\nu h_{\mu\nu}\partial^\rho 
h_{\mu\rho}-\frac{1}{2}\partial_\mu h\partial^\nu h_{\mu\nu}\right)+\\
&\nonumber~~~~~~~~+\lambda\dot{x}^\mu\left(-\frac{h^2}{2}p_\mu-h_{\mu\nu}^2p_\mu
+2hh_{\mu\nu}p^\nu\right)+i\dot{x}^\mu\left(-2(\partial^\rho
 h_{\mu\nu}) h_{\nu\rho}+h_{\rho\nu}\partial_\mu h_{\rho\nu}\right)\\
&~~~~~~~~+\lambda\dot{x}^\mu\dot{x}^\nu\left(-\left(\frac{h^2}{2}+h_{\mu\nu}^2\right)
\eta_{\mu\nu}+hh_{\mu\nu}\right)\Bigg]
~.
\label{action}
\end{align} 
Note that in \cref{action} 
we have introduced  a bookkeeping 
parameter $\lambda$  and rescaled $p\to\lambda p$,  $t\to 
t/\lambda$ and 
$\kappa\to\kappa/\lambda$. This choice is convenient in order to perform the 
soft expansion as a Laurent expansion in $1/\lambda$. 

Eq.(\ref{ft}) can be elaborated further by following a series of 
manipulations first   
described in \cite{Laenen:2008gt}. Indeed, since the dressed propagator 
$\bra{p_f}(2i(H-i\epsilon))^{-1}\ket{x_i}$ in the on-shell limit develops a 
simple pole with residue one and because the factor $f(T)$ remains finite in 
this 
limit, we can integrate by parts to get
\begin{align}
(p_f^2-m^2+i\epsilon) 
\,\bra{p_f}(2i(H-i\epsilon))^{-1}\ket{x_i}
&=e^{ip_fx_i}
\int_0^{\infty} dT\,\left(
\frac{d}{dT}e^{-\frac{i}{2}(p_f^2-m^2)T}
\right)
\, e^{-\epsilon T} f(T) \notag \\
=&e^{ip_fx_i}\left(
-f(0)
-\int_0^{\infty} dT\,
e^{-\frac{i}{2}(p_f^2-m^2)T}
\,\frac{d}{dT} \left(e^{-\epsilon T} f(T) \right)\right)\notag \\ 
=&e^{ip_fx_i}
\lim_{T\to\infty}
e^{-\epsilon T} f(T) ~,
\label{ft2}
\end{align}
where in the last step we took the on-shell limit $p_f^2\to m^2$.  
Therefore, \cref{ft} reads 
\begin{equation}
(p_f^2-m^2+i\epsilon) 
\,\bra{p_f}(2i(H-i\epsilon))^{-1}\ket{x_i}=
e^{ip_fx_i}\int\limits_{\tilde 
	x(0)=0}\mathcal{D}\tilde x 
\exp\left(i\int\limits_{0}^{\infty}dt~e^{-\epsilon t} L[\tilde x(t)]\right)~.
\label{finfty}
\end{equation}

A few comments are in order. Since the graviton field is a function of the 
spacetime variable $x^{\mu}(t)=x_i+p_ft+\tilde x(t)$, 
the Lagrangian in \cref{action} and \cref{finfty} defines a one-dimensional 
quantum field theory 
for 
the field $\tilde x^{\mu}(t)$, representing the fluctuations in the trajectory 
of the 
scalar particle due to graviton emissions. 
As first noted in \cite{Laenen:2008gt}, it is convenient to evaluate the path 
integral for $x_i=0$. The effect of having $x_i\neq 0$ is computed separately 
and, after combining it with the 
internal emission discussed in \cref{sec:fact}, yields a combination of 
derivatives of the non-radiative process, which corresponds to a coupling with 
the orbital angular momentum of the scalar particle.

Path integrals like in \cref{finfty} are typically solved in the worldline 
formalism at the \emph{amplitude} level, i.e. after combining all dressed 
propagators of the amplitude. In this way one obtains master formulae 
that generate integral representations of the amplitude order by order in the 
coupling constant $\kappa$, corresponding to a fixed number of graviton 
emissions. The GWL approach is different. In the gauge theory case the GWL is 
constructed from the evaluation of 
the equivalent expression to \cref{finfty} for each asymptotic particle, order 
by order in the soft 
expansion (i.e. assuming the soft momentum $k\ll p_f$) but considering an 
arbitrary 
number of soft emissions. In this way, the worldline path integral 
is solved once for each external scalar line, while the amplitude is 
constructed in terms of vacuum expectation values of GWLs using the action 
governing the background field. 
The same approach can be implemented in gravity, with some complication due to 
the presence of the weak field 
expansion. In fact, while one has still $k\ll p_f$, higher orders in $\kappa$ 
are also suppressed, which can be seen as a consequence of the charge being the 
four-momentum.

To avoid ambiguities between the weak and soft expansions, in \cref{action}
we have introduced  a single bookkeeping 
parameter $\lambda$ and rescaled $p\to\lambda p$,  $t\to 
t/\lambda$ and 
$\kappa\to\kappa/\lambda$. In this way, the path integral is evaluated order by 
order in $1/\lambda$. To see why this is possible, one can observe 
 that 
 expanding the background field 
 $h^{\mu\nu}(p_ft+ x)$ around $p_ft$ generates 
 powers of $ x$ with no 
 $\lambda$ enhancement. 
On the other hand, 
two-point correlators of 
$x$  are of order $1/\lambda$, and are given by
\begin{align}
\contraction{}{x_\mu(t)}{}{x_\nu(t')}
x_\mu(t)x_\nu(t')&=\frac{-i}{\lambda}\min(t,t')\eta_{\mu\nu}~,
\label{corr1}\\
\contraction{}{\dot{x}_\mu(t)}{}{x_\nu(t')}
\dot{x}_\mu(t)x_\nu(t')&=\frac{-i}{\lambda}\theta(t'-t)\eta_{\mu\nu}~,
\label{corr2}\\
\contraction{}{\dot{x}_\mu(t)}{}{\dot{x}_\nu(t')}
\dot{x}_\mu(t)\dot{x}_\nu(t')&=\frac{-i}{\lambda}\delta(t'-t)\eta_{\mu\nu}~.
\label{corr3}
\end{align}
Therefore, only a finite number of diagrams is necessary at a given order in 
$1/\lambda$.
At this point, we note that
equal-time correlators are ill-defined. The prescription 
used in \cite{White:2011yy} is to set to zero both 
$\theta(t'-t)$ and 
$\delta(t'-t)$ at equal time. 
In the following section we will discuss how this choice can be justified.

Although the 
weak field expansion differs from the one of 
\cite{White:2011yy}, the calculation of the path integral in \cref{finfty} is 
essentially the same. However, the 
number of terms is greatly reduced. For the sake of completeness an explicit 
evaluation of 
the worldline diagrams is discussed in \cref{sec:diag}. 
The upshot is that the sum of the connected diagrams is equal to the sum
of the terms in \cref{Evertex}, 
\cref{NEvertex} and \cref{NEvertices}.
It is worth mentioning that, as in 
\cite{White:2011yy}, it
is still necessary to expand the Hamiltonian up to order $\kappa^2$ in order to 
correctly reproduce these vertices. Thus, we conclude that this 
feature is not due to the particular definition of the graviton field in 
\cref{weak} or 
\cref{weakg}, but it is due to the fact that the 
Hamiltonian has been defined after a weak field expansion. 

The final step consists of exponentiating the result obtained 
from the 
worldline diagrams. In fact, remembering the standard property valid for any 
QFT with commuting sources that 
the sum of connected 
diagrams exponentiates, one obtaines 
\begin{align}
(p_f^2-m^2+i\epsilon) &
\,\bra{p_f}(2i(H-i\epsilon))^{-1}\ket{0}
=\exp\Bigg(\kappa\int \frac{d^dk}{(2\pi)^d}
\left(V_{\mu\nu}^{\text{E}}(k) 
+V_{\mu\nu}^{\text{NE}}(k) \right)\tilde h^{\mu\nu}(k)
\notag \\
&
\qquad \qquad +\kappa^2\int \frac{d^dk}{(2\pi)^d}\int \frac{d^dl}{(2\pi)^d}
V_{\mu\nu\rho\sigma}^{\text{NE}}(k,l) \tilde h^{\mu\nu}(k)\tilde 
h^{\rho\sigma}(l)
\Bigg)~,
\label{gwlfourier}
\end{align}
where $\tilde h^{\mu\nu}$ denotes the Fourier transform of the graviton field, 
while $V_{\mu\nu}^{\text{E}}$, $V_{\mu\nu}^{\text{NE}}$ and 
$V_{\mu\nu\rho\sigma}^{\text{NE}}$  have been defined in \cref{Evertex}, 
\cref{NEvertex} and \cref{NEvertices}. This proves the exponentiation of the 
next-to-soft vertices discussed in \cref{sec:fact}, thus showing that the 
next-to-soft factorization of \cref{NEid} persists to all orders in $\kappa$. 
Finally, performing the inverse Fourier transform in the r.h.s. of
\cref{gwlfourier}   
one obtains the GWL in position space defined in \cref{gwl}.

\section{GWL from worldline model in curved space}
\label{sec:worldline}

One of the crucial ingredients of 
the derivation outlined in the previous section is the definition of 
the Hamiltonian in \cref{eq:Hamiltonian}, which is obtained \emph{after} a weak 
field 
expansion in the field theory Lagrangian. This approach is not satisfactory for 
a number 
of reasons. The first one is that it runs into difficulties when one 
attempts to generalize the definition to spinning emitters, as already observed 
in the gauge theory case in \cite{Bonocore:2020xuj}. 
Secondly, the need to expand the Hamiltonian to order $\kappa^2$ in order to 
reproduce the vertices in \cref{NEvertices} is somewhat unconventional compared 
to the standard worldline formulation where the two-graviton vertex is 
reproduced by the order $\kappa$ terms in the Hamiltonian. 
Finally (and perhaps more importantly), 
the above construction hides non-trivial cancellations of UV divergences, which 
arise from the equal-time delta function in \cref{corr3}. This issue has 
been thoroughly discussed in the literature, and it 
requires the use of ghost 
fields and regularization scheme dependent
counterterms in the Hamiltonian. It is not clear from the previous construction 
how the Hamiltonian in \cref{eq:Hamiltonian} is affected by the UV 
regularization and what the role of the ghost fields is in the evaluation of 
the exponentiated worldline diagrams. 
In this section, we will demonstrate that a 
first-principles derivation of the GWL in the worldline formalism and 
the identification of the correct Hamiltonian before any weak 
field expansion   
clarify the above issues. In this way the construction of the GWL is put on a 
firm 
basis.

\subsection{Quantization in curved space}
\label{sec:quant}

The quantization of a relativistic particle on curved spacetime has a long 
history\footnote{For a comprehensive review see \cite{Bastianelli:2006rx} and 
references therein.}.
Most of the difficulties revolve around the issue of a unique definition of an 
Hamiltonian operator, due to the presence of UV 
divergences in the worldline model and the lack of an unambiguous definition
for the momentum operator in absence of translational invariance 
\cite{DeBoer:1995hv}.

The reason for the presence of UV 
divergences is relatively simple. A particle in 
flat spacetime can be described by the following classical phase space action
\begin{equation}
S[x,p]=\int dt\, \left(-p_{\mu} \dot x^{\mu}+e H(x,p) \right)~,
\label{polyakov}
\end{equation}
where the einbein $e$ is the Lagrange multiplier for the constraint $H(x,p)=0$.
After quantizing the constrained model \`a la Dirac, and gauge fixing $e=T$,
one obtains the usual Schwinger representation of the dressed propagator as in 
\cref{schwinger}. With most Hamiltonians the momentum 
can be integrated out exactly, so that the action in configuration space
takes the form
\begin{align}
S[x]=	\int\limits_{0}^{T}dt \left(-\frac{1}{2}\eta_{\mu\nu}\dot x^{\mu} \dot 
x^{\nu}+V(x) 
	\right)
~,
\end{align}
where $V(x)$ can be derived from the Hamiltonian. 
When moving to curved space,
the kinetic term becomes a non-linear sigma model 
of the type $g_{\mu\nu}(x)\dot x^{\mu} \dot x^{\nu}$. Although 
superrenormalizable, some diagrams might contain UV divergences\footnote{The 
presence of IR divergences in the worldline path integral is instead more 
subtle. They are typically regulated by the proper time $T$. However, in the 
asymptotic limit one eventually takes the limit $T\to\infty$. This has the 
consequence that IR divergences show up in the vacuum expectation values of 
multiple GWLs, as expected.}. More specifically, the divergences come from 
correlators of $\dot x(t)$ evaluated at equal time, a feature evident in 
\cref{corr3} in the previous section and also in four-dimensional 
theories \cite{Gerstein:1971fm}. 

As it is well-known \cite{Bastianelli:1992ct}, to solve the problem one needs 
to define a regularization prescription and to subsequently 
renormalize the divergences with suitable counterterms, with a scheme dependent 
finite remainder 
in this cancellation.
Among the various regularization schemes, time slicing has the advantage that 
such cancellation is already built-in in the path integral measure. Indeed, 
integrating 
out the momentum yields a factor $(-g(x))^{-1/2}$ that can 
be exponentiated in terms of ghost fields, whose correlators should cancel the 
divergences in $\langle\dot x(t)\dot x(t) \rangle$.
Unlike in other methods, such as mode 
regularization 
or dimensional regularization, no counterterm must be inserted by hand, since 
the additional term in the 
Hamiltonian operator $\hat H$ naturally arises from ordering the operators 
$\hat x$ and $\hat p$. 

Clearly, discussing the Hamiltonian ordering assumes that a suitable  
definition for $\hat H$ has been found within a meaningful quantization 
procedure, as we now briefly review. 
We start by observing that the  
 constraint $\hat 
 H|\psi\rangle=0$ should be equivalent to the 
Klein-Gordon equation 
\begin{align}
(\nabla_{\mu}g^{\mu\nu}\partial_{\nu}+m^2)\phi(x)=0~,
\label{KG}
\end{align}
where $\nabla_{\mu}$ denotes the covariant derivative. 
In this way, $\hat H$ contains the covariant Laplacian and it is automatically 
 invariant under general relativity transformations. 
Once a definition for the momentum operator $\hat p$
in position representation is adopted, the Hamiltonian operator $\hat H$ 
follows 
unambiguously.
A common choice in the literature is given by 
\begin{align}
\langle x|\hat p_{\mu}|\psi\rangle&=(-g(\hat 
x))^{-1/4}(i\partial_{\mu})(-g(\hat x))^{1/4}
\,\langle x|\psi\rangle~, 
\label{herm}
\end{align}
which yields
\begin{align}
2\hat H&=-(-g(\hat x))^{-1/4}\,\hat  p_{\mu}\,g^{\mu\nu}(\hat x)\,
\sqrt{-g(\hat x)}\,\hat 
p_{\nu}\,(-g(\hat x))^{-1/4}+m^2~, 
\label{Hherm}
\end{align}
where the $\hat x$ dependence has been made explicit. 
One can readily verify that the operator in \cref{herm} is hermitian w.r.t. 
the Hilbert space inner product, defined as
\begin{align}
\langle\psi|\chi\rangle&=
\int d^dx \, \sqrt{-g}\, \psi^*(x)\chi(x)~,
\label{inner}
\end{align}
where we normalized the $\hat x$ eigenstates via
\begin{align}
1=\int d^dx\,\sqrt{-g}\, |x\rangle\langle x|~, \qquad 
\langle x|x'\rangle=\frac{1}{\sqrt{-g}}\delta(x-x')~.
\label{compl}
\end{align}
Note that
the definition in \cref{herm} is consistent with the normalization 
$\langle x|p 
\rangle=\langle p|x 
\rangle^{\dag}=(-g)^{-1/4}e^{-ip_{\mu}x^{\mu}}$. 
This is particularly convenient in Weyl ordering, i.e. 
when symmetrizing all $\hat x$ and $\hat p$ operators.

In the
construction of the GWL, on the other hand, the building block is the 
asymptotic propagator 
$\langle p_f|e^{-iHt}|x_i \rangle$, as we discussed in the previous section. 
Hence, even if we consider the position space dressed propagator, some kind of 
Fourier transform must be defined before taking the weak field expansion. 
In fact,
although the dependence over the conjugate momentum $p$ is Gaussian and the 
non-trivial path integral is the one over $x$, the latter is evaluated 
pertubatively in the soft 
expansion by adding fluctuations over the straight classical path in 
\emph{flat} spacetime, via the scaling defined by the final momentum $p_f$. 
This asymmetry between initial and final state makes $px$-ordering preferable 
w.r.t Weyl ordering.

These considerations lead us to define the Fourier transform as \footnote{This 
definition has been 
 extensively discussed in 
\cite{Horwitz:2020cde}.}
\begin{align}
\tilde\psi(p)&\equiv \langle p|\psi\rangle=\int 
d^dx\,\sqrt{-g}\,\psi(x)\,e^{ip_{\mu}x^{\mu}},
\label{fourier}
\end{align}
or equivalently we define the left eigenstates of $\hat p$ via $\langle p|x 
\rangle=e^{ip_{\mu}x^{\mu}}$. 
Using the normalization of 
\cref{compl}, it is easy to see that \cref{fourier} implies
\begin{align}
\langle x| \hat p_{\mu}|\psi\rangle &=(-g(\hat 
x))^{-1/2}(i\partial_{\mu})(-g(\hat x))^{1/2}\, \langle x|\psi\rangle~. 
\label{nonherm}
\end{align}
This definition is not hermitian w.r.t. the inner 
product of \cref{inner}. 
Note also that $\langle x|p 
\rangle=(-g)^{-1/2}e^{-ip_{\mu}x^{\mu}}\neq \langle p|x 
\rangle^{\dag}$ and that
 the 
wavefunction $\tilde\psi(p)$ in \cref{fourier} is not a scalar in general 
relativity.  
However, the Hamiltonian must be invariant. In fact, by demanding that 
$\hat H|\psi\rangle=0$ returns \cref{KG} with the momentum in 
\cref{nonherm}, 
the Hamiltonian is uniquely defined and reads 
\begin{align}
 2\hat H&=-\hat  p_{\mu}\,g^{\mu\nu}(\hat x)\,(-g(\hat x))^{1/2}\,\hat 
p_{\nu}\,(-g(\hat x))^{-1/2}+m^2~. 
\label{Hnonherm}
\end{align}

The non-hermitian result\footnote{We stress that both the operator 
	$\hat x$ and the Hamiltonian in \cref{Hnonherm} are hermitian.} in 
	\cref{nonherm} (or equivalently 
 the non-invariant Fourier transform defined in \cref{fourier}) 
might seem puzzling at first sight but it is a direct consequence of 
$px$-ordering\cite{Berezin:1971jf}. In fact, it 
does not pose any real problem, since
the physical momentum contained in the GWL is the momentum $p_f$ defined in 
flat spacetime and not the conjugate 
momentum $\hat p$.  The reason for this choice
can be appreciated only after defining the classical Hamiltonian $H_{px}(p,x)$ 
from   
\cref{Hnonherm} in $px$-ordering, i.e. 
\begin{align}
2H_{px}(p,x)&\equiv  -p_{\mu}\, 
p_{\nu}\,g^{\mu\nu}+m^2 
+ip_{\mu}(\partial_{\nu}g^{\mu\nu}+g^{\mu\nu}(\partial_{\nu}\ln(\sqrt{-g})))~. 
\label{xp}
\end{align}
The logarithmic term in \cref{xp} contains the trace of the metric and cannot 
be generated in $px$-ordering with 
the hermitian definitions of \cref{herm} and \cref{Hherm}, as one can readily 
verify. This term is crucial in order to correctly reproduce the two-graviton 
vertex, as we are going to discuss in \cref{sec:exp}. Before doing that, we 
need to 
set up a path integral representation for the asymptotic propagator.

\subsection{Setting up the path integral}

Equipped with the Hamiltonian in \cref{xp}, we can now work out a path integral 
representation for the dressed propagator in time slicing regularization. In 
analogy with the discussion in 
\cref{sec:review}, we consider 
\begin{align}
&\bra{p_f}e^{-i(\hat H-i\epsilon)t}\ket{x_i}=\notag\\
=&\int\limits \prod_{i=2}^{N}d^4x_i 
\sqrt{-g(x_i)}\prod_{j=1}^{N-1}\frac{d^4p_i}{(2\pi)^4}\prod_{n=1}^{N}\exp
\left(-iH_{px}(p_n,x_n)\tau \right)\braket{p_n|x_n}\prod_{m=1}^{N-1}
\braket{x_{m+1}|p_m}\notag \\
=&e^{ip_Nx_N}\int\limits 
\prod_{i=2}^{N}d^4x_i\prod_{j=1}^{N-1}\frac{d^4p_i}{(2\pi)^4}\exp
\left\{-i\sum_{n=1}^{N-1}p_n(x_{n+1}-x_n)-i\sum_{n=1}^{N}
H_{px}(p_n,x_n)\tau-i\epsilon\right\}~,
\label{discrete}
\end{align}
where we sliced the time domain in $N$ intervals of length 
$\tau=T/N$.
Note that \cref{discrete} is consistent with the Fourier transform defined 
in \cref{fourier}.

In order to carry out the Gaussian momentum integration, it is convenient to
 define
\begin{align}
V^{\mu}_n&\equiv 
\partial_{\nu}g^{\mu\nu}(x_n)+g^{\mu\nu}(x_n)(\partial_{\nu}\ln(\sqrt{-g(x_n)}))~,
\label{counter}
 \\
B^\mu_n&\equiv -i\tau 
\left[\frac{(x_{n+1}-x_n)^\mu}{\tau}+\frac{i}{2} V_n^\mu 
\right]~.
\end{align}
Then, \cref{discrete} becomes
\begin{equation}
\bra{p_f}e^{-i(\hat H-i\epsilon)t}\ket{x_i}=
e^{ip_Nx_N-\frac{i}{2}m^2T}\int\limits 
\prod_{i=2}^{N}d^4x_i 
\left(\prod_{j=1}^{N-1}\sqrt{\frac{g(x_j)}{(2\pi\tau)^n}}\right)
\exp\left\{\frac{i}{2\tau}\sum_{n=1}^{N-1}
B^\mu_n g_{\mu\nu}^n B^\nu_n\right\}.
\end{equation}
As anticipated in \cref{sec:quant}, we can get rid of the determinants in the 
$x$-measure 
by 
re-exponentiating them using three ghost fields, to get
\begin{equation}\label{Zdiscr}
\bra{p_f}e^{-i(\hat H-i\epsilon)t}\ket{x_i}=
e^{ip_Nx_N-\frac{i}{2}m^2T}\int\limits 
\prod_{i=2}^{N}\frac{d^4x_i}{(2\pi\tau)^n} 
\exp\left\{\frac{i}{2\tau}\sum_{n=1}^{N-1}g_{\mu\nu}^n
B^\mu_n B^\nu_n\right\}\cdot Z_{ghosts}~,
\end{equation} 
where $Z_{ghosts}$ reads
\begin{equation}\label{Zghosts}
Z_{ghosts}=\int\limits \prod_{i=1}^{N-1}d^4a_id^4b_id^4c_i 
\exp\left\{-i\frac{\tau}{2}\sum_{n=1}^{N-1}g_{\mu\nu}^n(a_n^\mu 
a_n^\nu+b^\mu_nc^\nu_n)\right\},
\end{equation}
where as usual we have neglected irrelevant field-independent normalization 
constants. 
Here, $a_i$ are real commuting 
fields, while $b_i$ and $c_i$ are anti-commuting 
Grassmann fields.

The continuum limit of \cref{Zdiscr} and \cref{Zghosts} follows 
straightforwardly. Expanding around the 
classical path
$x(t)=x_i+p_f\,t+\tilde x(t)$
and truncating the external free propagator,
we obtain a path integral representation for the asymptotic propagator in 
analogy with  
\cref{finfty}, i.e. 
\begin{equation}
(p_f^2-m^2+i\epsilon) \,
\bra{p_f}(2i(H-i\epsilon))^{-1}\ket{x_i}
=e^{ip_fx_i}\int\limits_{\tilde 
x(0)=0}\mathcal{D}\tilde x 
\mathcal{D}a \mathcal{D}b \mathcal{D}c
\exp\left(i\int\limits_{0}^{\infty}dt~e^{-\epsilon t} ~L[\tilde 
x,a,b,c]\right)~,
\label{finfty2}
\end{equation}
where the Lagrangian $L[\tilde x,a,b,c]$ reads
\begin{align}
L[\tilde x,a,b,c]=- 
\frac{1}{2}
\left(
(\dot{\tilde x}^{\mu}\dot {\tilde x}^{\nu}
+  {a}^{\mu}  {a}^{\nu}
+  {b}^{\mu}  {c}^{\nu})
g_{\mu\nu}
+i(\dot {\tilde x}+p_f)^{\mu}g_{\mu\nu}V^{\nu}
-\frac{1}{4}V^{\mu}g_{\mu\nu}V^{\nu}
\right)~.
\label{asympt-exact}
\end{align}
We stress that although we have expressed the path integral in terms of the 
classical straight solution in flat space, so far we have not performed any 
weak 
field expansion. Hence,   
the expression in \cref{asympt-exact} holds for a generic curved 
background, as long as the Fourier transform of \cref{fourier} is meaningful. 

At this point one would like to solve the path integral perturbatively, hence 
the need for the weak field expansion of \cref{weak}. While vertices can be 
read immediately from \cref{asympt-exact}, 
two-point correlators for all fields must be derived from the discrete version 
in \cref{Zdiscr} and \cref{Zghosts} in order to avoid ambiguities at 
equal-time. Although similar 
calculations have been thoroughly discussed in 
the literature (see e.g. \cite{Bastianelli:2006rx}), 
given the non-standard 
conventions 
required by the GWL (i.e. $px$-ordering and the asymptotic limit), and the 
quite laborious algebra, we present the corresponding derivation in 
\cref{sec:corr}. 
The upshot is that the propagators for the ghost fields read
\begin{align}
\langle a^{\mu}(t)a^{\nu}(t')\rangle&=-i\eta^{\mu\nu}\delta(t-t')~,
\label{ghost1}\\
\langle b^{\mu}(t)c^{\nu}(t')\rangle&=2i\eta^{\mu\nu}\delta(t-t')~,
\label{ghost2}
\end{align}
while
 the two-point correlators for the $x^{\mu}$ fields assume exactly 
the same form 
as in \cref{corr1}, \cref{corr2} and \cref{corr3}. This time, however, their 
equal-time expression is 
non-ambiguous and follows from the regularization scheme. In particular, 
we obtain  
\begin{align}
\langle \dot x^{\mu}(t)x^{\nu}(t)\rangle & =0~, \label{corrdot1}\\
\langle 
\dot x^{\mu}(t)\dot 
x^{\nu}(t)\rangle&=-i\eta^{\mu\nu}\delta(0)~,\label{corrdot2}\\
\langle a^{\mu}(t)a^{\nu}(t)\rangle&=-i\eta^{\mu\nu}\delta(0)~,\label{corra}\\
\langle b^{\mu}(t)c^{\nu}(t)\rangle&=2i\eta^{\mu\nu}\delta(0)~.\label{corrbc}
\end{align}
This simple result
shows that the correlator $\langle 
\dot x^{\mu}(t)\dot 
x^{\nu}(t)\rangle$ is divergent in time slicing, but that the divergence is 
compensated by the ghost correlators $\langle a^{\mu}(t)a^{\nu}(t)\rangle$
and $\langle b^{\mu}(t)c^{\nu}(t)\rangle$, in agreement with similar 
calculations in Weyl-ordering.
On the other hand, unlike in Weyl-ordering, we do not obtain the midpoint rule  
$\langle \dot 
x^{\mu}(t)x^{\nu}(t)\rangle=\theta(0)\,\eta^{\mu\nu}=1/2\,\eta^{\mu\nu}$ 
\footnote{
This difference can be noted also in the gauge theory case discussed in 
\cite{Laenen:2008gt} and \cite{Bonocore:2020xuj}. There, one uses $\theta(0)=0$ 
while the next-to-soft vertex $k^{\mu}/2p\cdot k$, analogous to \cref{NEvertex} 
in this paper, is due to the term 
$\partial_{\mu}A^{\mu}$ in the $px$-ordered Hamiltonian. This term 
is absent in Weyl-ordering, but the equal-time correlator $\langle\dot x(t) 
x(t)\rangle$ is non-vanishing and compensates for the mismatch.   
}. 
However, one should not be fooled by the simplicity of the results of 
 \crefrange{corrdot1}{corrbc} which hide 
some non-trivial manipulations in the discrete case, as shown in 
\cref{sec:corr} and \cref{sec:trig}.

The soft expansion is carried out with the same manipulations as in 
\cref{sec:review}. After rescaling $p\to\lambda p$,  $t\to 
t/\lambda$ and 
$\kappa\to\kappa/\lambda$, all two-point correlators of $x$ are of order 
$1/\lambda$ 
while powers of $x$ originating from the Taylor expansion of $h^{\mu\nu}(x)$ 
contain no $\lambda$ enhancement.
For what concerns 
the ghost fields instead, it is convenient to rescale them via
$\{a,b,c\} \to  \{\lambda a,\lambda b, \lambda c\}$.
Then, we observe that
the inverse metric (hence corrections of order  $\kappa^2$) appears in 
the exponent of \cref{asympt-exact} only via $V^{\mu}$. Therefore, 
at order $1/\lambda$ we can just drop ${\cal O}(\kappa^2)$ in the Hamiltonian 
as well as the term quadratic in $V^{\mu}$ in \cref{asympt-exact}. 
This is one of the major differences and advantages of this derivation of the 
GWL compared to 
the derivation of \cref{sec:review}.
In fact,  
making the weak field expansion of \cref{weak} and powers of $\lambda$ 
explicit, 
the relevant terms in the Lagrangian of \cref{asympt-exact} for an evaluation 
of 
the path integral at order $1/\lambda$ reduce to
\begin{align}
L[\tilde x,a,b,c]&=-
	\frac{\lambda}{2}\left(\dot{\tilde x}^2+{ a}^2+ b_{\mu} c^{\mu}\right)
	-\frac{\kappa}{2}(a^{\mu}a^{\nu}+b^{\mu}c^{\nu})h_{\mu\nu}
	-\kappa p_f^{\mu}\dot {\tilde x}^{\nu}h_{\mu\nu}
	- \frac{\kappa}{2}p_f^{\mu}p_f^{\nu}h_{\mu\nu}
		\notag \\
	& \qquad 
	- \frac{\kappa}{2}  \dot {\tilde x}^{\mu} \dot {\tilde x}^{\nu} h_{\mu\nu}
	-i\frac{\kappa}{2\lambda}
	p_f^{\mu}
	\left(
	\frac{1}{2}\partial_{\mu}h-\partial_{\nu}h_{\mu\nu}
	\right)
	~.
\label{asympt-expanded}
\end{align}
Despite the presence of ghost fields, one can appreciate the drastic 
simplification of \cref{asympt-expanded} w.r.t. \cref{action}, where the action 
had to be expanded up to order $\kappa^2$. We are now ready to solve the path 
integral.

\subsection{Worldline exponentiation}
\label{sec:exp}

As already mentioned in \cref{sec:review}, it is convenient to evaluate the 
path integral for $x_i=0$, since the effect of having $x_i\neq 0$
combines with internal emissions to give the orbital angular momentum 
$L^{\mu\nu}$ of the scalar particle. 
Then, 
the calculation is straightforward and boils down to inserting the expansion
\begin{align}
h_{\mu\nu}(pt+x)=h_{\mu\nu}(pt)
+x^{\rho}\partial_{\rho}h_{\mu\nu}(pt)
+\frac{1}{2}x^{\rho}x^{\sigma}\partial_{\rho}\partial_{\sigma}h_{\mu\nu}(pt)
+{\cal O}(x^3)
\end{align}
into \cref{asympt-expanded}, where 
for the brevity of the notation we replaced $\tilde 
x\to x$ and $p_f\to p$. 
The relevant vertices at order $1/\lambda$ are
\begin{align}
\textcircled{1}&=-\frac{i\kappa}{2}p_\mu p_\nu  \int\limits_{0}^{\infty}dt \,
x^{\rho} \partial_\rho h^{\mu\nu}(pt) \label{v1}~,\\
\textcircled{2}&=-i\kappa  p^\nu \int\limits_{0}^{\infty}dt\,
\dot{x}^\mu\,h_{\mu\nu}(pt)\label{v2}~,
\end{align}
\begin{align}
\textcircled{3}&=-\frac{i\kappa}{2} p_\mu p_\nu \int\limits_{0}^{\infty}dt 
x^\rho x^\sigma\partial_\rho \partial_\sigma h^{\mu\nu}(pt) \label{v3} ~, \\
\textcircled{4}&=-\frac{i\kappa}{2} 
\int\limits_{0}^{\infty}dt\, \dot {\tilde x}^{\mu} \dot {\tilde x}^{\nu} 
h_{\mu\nu}(pt) \label{v4}~,\\
\textcircled{5}&= 
-\frac{i\kappa}{2}(a^{\mu}a^{\nu}+b^{\mu}c^{\nu})h_{\mu\nu}(pt) 
\label{v5}~.
\end{align}
Moreover, for the evaluation of the path integral at order $1/\lambda$ one also
needs the following terms with no power of $x$:
\begin{align}
-\frac{i\kappa}{2}\int\limits_{0}^{\infty}dt p_\mu p_\nu h^{\mu\nu}(pt)
+\frac{\kappa}{2\lambda}
\int\limits_{0}^{\infty}dt\,{p}_{\mu}
\left(
\frac{1}{2}\partial^{\mu}h(pt)-\partial_{\nu}h^{\mu\nu}(pt)
\right)~.
\label{nox}
\end{align}
Note that all the above integrals are regulated at $t\to\infty$ by 
a factor $e^{-\epsilon t}$ which originates from 
the Feynman $+i \epsilon$ prescription in 
\cref{finfty2}. 

At order $\lambda^0$ we consider the path integral at its stationary point, 
without any propagating field. 
Thus we 
consider only the first term in \cref{nox}, which yields
\begin{align}
-\frac{i\kappa}{2}\int\limits_{0}^{\infty}dt p_\mu p_\nu h^{\mu\nu}
=\int\limits 
\frac{d^4k}{(2\pi)^4}\tilde{h}_{\mu\nu} \left[-\frac{\kappa}{2}\frac{p^\mu 
p^\nu}{pk}\right]~,
\label{ruleE}
\end{align}
in agreement with \cref{Evertex}.

At order $1/\lambda$ we start including (connected) Feynman diagrams.  
We distinguish contributions of order $\kappa$ and $\kappa^2$, respectively. 
We start by noting that vertices $\textcircled{4}$ and  $\textcircled{5}$, 
combined with the respective equal time propagators, cancel exactly.
This was anticipated given that the role of the ghost fields is to cancel the 
UV divergences of $\langle \dot x(t) \dot x(t)\rangle $. 
Then,
at order $\kappa$ we are left with the second term in \cref{nox} and
the combination of the vertex $\textcircled{3}$ with the equal time propagator 
$\langle x(t)  x(t)\rangle $. They yield
\begin{align}
&\nonumber\frac{i\kappa}{2}\int\limits_{0}^{\infty}dt~\Bigg(-\frac{1}{2} 
\partial_\rho\partial_\sigma h_{\mu\nu}p^\mu p^\nu \wick{\c x^\rho(t)\c 
x^\sigma(t)}+i(\partial^\mu h_{\mu\nu}p^\nu-\frac{1}{2}\partial_\mu 
hp^\mu)\Bigg)=\\
=&\int\limits\frac{d^4k}{(2\pi)^4}\tilde{h}_{\mu\nu}~\left[-\frac{\kappa}{4}\frac{k^2p^\mu
 p^\nu}{(pk)^2}+\frac{\kappa}{4}\frac{p^\mu k^\nu+p^\nu 
k^\mu-pk\eta^{\mu\nu}}{pk}\right]~,
\label{NEpath}
\end{align}
in agreement with \cref{NEvertex}.

At order $\kappa^2$ we consider the following contractions:
\begin{align}
\textcircled{1}-\textcircled{1}&=\int\limits 
\frac{d^4kd^4l}{(2\pi)^8}\frac{\tilde{h}_{\mu\nu}\tilde{h}_{\rho\sigma}}{2}\left[\frac{\kappa^2}{4}\frac{kl~p^\mu
 p^\nu p^\rho p^\sigma}{pl~pk~p(k+l)}\right]~,\label{rmunu1}\\
\textcircled{2}-\textcircled{2}&=\int\limits 
\frac{d^4kd^4l}{(2\pi)^8}\frac{\tilde{h}_{\mu\nu}\tilde{h}_{\rho\sigma}}{2}\left[\kappa^2
\frac{ \eta^{\mu\rho} p^\nu 
p^\sigma}{p(k+l)}\right]~,\label{rmunu2}\\
\textcircled{1}-\textcircled{2}&=\int\limits 
\frac{d^4kd^4l}{(2\pi)^8}\frac{\tilde{h}_{\mu\nu}\tilde{h}_{\rho\sigma}}{2}\left[-\kappa^2\frac{p^\mu
 p^\nu p^\sigma k^\rho}{kp~p(k+l)}\right]~.\label{rmunu3}
\end{align}
When symmetrizing the last two terms with respect to 
$\mu\leftrightarrow\nu,\rho\leftrightarrow\sigma$ and with respect to 
$(k,\mu\nu)\leftrightarrow (l,\rho\sigma)$ one ends up with 
\cref{NEvertices}.
Therefore, the sum of the diagrams at order $1/\lambda$ has returned the 
sum of the factorized vertices in \cref{Evertex}, \cref{NEvertex} and 
\cref{NEvertices}, in agreement with the result of \cref{sec:review}. 
Note that this is non-trivial, given the two different
 Hamiltonians in the respective calculations.
In particular, one can appreciate the relative simplicity of the calculation 
of the 
two-graviton vertex in this section by comparing \cref{rmunu1}, \cref{rmunu2} 
and \cref{rmunu3}  
with the equivalent
calculation of \cref{sec:review} outlined in \cref{sec:diag}. 

Finally, 
one can use once again the well-known theorem in QFT that states that 
connected 
diagrams 
exponentiate. Therefore, one obtains  the 
desired 
representation of the GWL, in agreement with the expression in 
Fourier space 
(\cref{gwlfourier}) and in position space (\cref{gwl}).

At this point, it is interesting to note that the worldline approach of this 
section makes clear the importance of the Fourier transform defined in 
\cref{fourier} 
within our regularization prescription.
More specifically, the term proportional to $\eta_{\mu\nu}$ in \cref{NEpath}
comes from the trace of $h_{\mu\nu}$, which in turn can be generated only by 
the logarithmic term in \cref{xp}. As we have discussed in \cref{sec:quant}, 
the presence of this 
term requires a non-hermitian momentum $\hat p$, which is a consequence of the 
definition of \cref{fourier}.

\section{Amplitude level and classical limit}
\label{sec:ampl}
In the previous section we have derived the GWL by solving the worldline model 
in curved space  
for each external leg of a scattering amplitude. This generalizes the 
well-known 
exponentiation of the background field in terms of Wilson lines, and thus  
forms \emph{per se} an 
exponentiation, that we can dub \emph{worldline} exponentiation. 
Of course, this is not the end of the story, since we have to combine several 
GWLs to extract information about physical observables.
In fact, GWLs are only the first step 
towards two other 
exponentiations at the amplitude level, i.e. the \emph{eikonal} and the 
\emph{soft} ones, obtained by taking the Vacuum Expectation Value (VEV) of GWLs 
using the 
(gauge fixed) Einstein-Hilbert action. In the next section we briefly review 
these exponentiations.

\subsection{Soft and eikonal exponentiation}
It has been known for a long time that 
Wilson lines provide a convenient representation of scattering amplitudes.
In particular, there are two different 
set-ups where Wilson lines are 
particular 
convenient to show all-order properties: the 
factorization of soft divergences and the Regge 
limit. Let us briefly review them in turn.

\begin{figure}
	\centering
	\includegraphics[width=100mm]{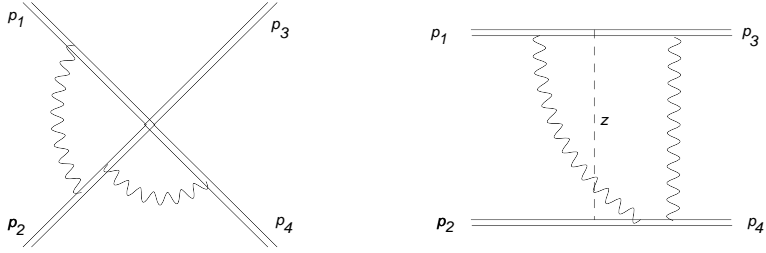}
	\caption{Sample diagrams for the VEV in \cref{eq:softfunct} 
		(left) and \cref{eq:regge2} (right). Wilson lines and gravitons are 
		denoted with 
		double 
		lines and wavy lines, respectively. The dashed line in the right
		diagram splits 
		each Wilson line in two 
		branches to 
		distinguish  
		initial and final states of the 
		leading order amplitude, where		
		two highly energetic particles are separated by an impact parameter $z$.
}
	\label{fig:soft-regge}
\end{figure}

Soft divergences in a $n$-point scattering amplitude factorize as\footnote{Here 
	we focus 
	on soft divergences and 
	neglect all other singularities.} 
${\cal A}_n= {\cal S}_n\times {\cal H}_n$, 
where ${\cal S}_n$ is a universal soft function given by the 
VEV of the time-ordered-product of straight semi-infinite 
Wilson 
lines $W$ originating from a point-like hard interaction ${\cal H}_n$. 
For instance, as shown in \cref{fig:soft-regge}, for a 
$2\to2$ process the soft function reads 
\begin{align}
{\cal S}&=\langle 0 |W_{p_1}(-\infty,0)
W_{p_2}(-\infty,0)W_{p_3}(0,\infty)W_{p_4}(0,\infty)|0\rangle~,
\label{eq:softfunct}
\end{align}
where the gravitational Wilson line on the direction $p$ has been defined as
\begin{align}
W_p(\lambda_1,\lambda_2)&\equiv\exp\left(
- \frac{i\kappa}{2}\int_{\lambda_1}^{\lambda_2} d\lambda\, 
p^{\mu}p^{\nu}h_{\mu\nu}(\lambda p)
\right)~.
\label{wilson}
\end{align}
One can take a step further and rearrange the product in \cref{eq:softfunct} 
into a single 
exponential ${\cal S}=e^{i{\cal W}}$. This is quite straightforward in 
gravity, where the exponent ${\cal W}$ is 
simply given by the one-loop result.
Things are more subtle in gauge theories, where the 
diagrams 
contributing to ${\cal W}$ have a richer structure and go under the name of 
\emph{webs} 
\cite{Yennie:1961ad, Gatheral:1983cz, Frenkel:1984pz, Mitov:2010rp, 
Gardi:2010rn, Gardi:2013ita, Falcioni:2014pka, 
	White:2015wha, Hannesdottir:2019opa}.

An independent exponentiation 
\cite{White:2019ggo} occurs in the study of the high energy limit of 
four-point scattering amplitudes (also called Regge limit), i.e. 
$2\to2$ processes in the limit where the center of mass energy $\sqrt{s}$ is 
much 
larger 
than the momentum transfer $\sqrt{|t|}$. This approach is rooted in the
study of 
perturbative quantum 
gravity at transplanckian energies, and recently attracted a renewed 
attention due to its connection to the classical regime. 
Given the highly forward limit of the process, the outgoing particles 
essentially do not recoil and interact with a very low energetic graviton. 
Hence, Wilson lines 
provide again a convenient and elegant formalism, as originally 
proposed by 
\cite{Korchemskaya:1994qp, 
	Korchemskaya:1996je, Melville:2013qca}. In fact, in analogy with 
\cref{eq:softfunct} and as shown in \cref{fig:soft-regge}, one can factorize 
the leading order amplitude ${\cal 
	A}_{\text{LO}}$ in 
the full amplitude ${\cal A}$
in terms of a VEV of four semi-infinite Wilson lines along the directions of 
the 
incoming and outgoing 
particles, separated by a (large) distance $z$, i.e.
\begin{align}
{\cal A}&={\cal A}_{\text{E}} \times {\cal 
	A}_{\text{LO}}~,
\label{eq:regge}
\end{align}
where the eikonal function ${\cal A}_{\text{E}}$ reads
\begin{align}
{\cal A}_{\text{E}}&=\langle 0 
|W_{p_1}(0,-\infty,0)
W_{p_2}(z,-\infty,0)
W_{p_3}(0,0,\infty)
W_{p_4}(z,0,\infty)
|0\rangle ~.
\label{eq:regge2}
\end{align}
Here, we had to include the possibility of a constant off-set $z$ by 
defining 
the 
Wilson line as
\begin{align}
W_p(z,\lambda_1,\lambda_2)&=\exp\left(
- \frac{i\kappa}{2}\int_{\lambda_1}^{\lambda_2} d\lambda\, 
p^{\mu}p^{\nu}h_{\mu\nu}(\lambda p+z)
\right)~.
\label{wilsonZ}
\end{align}
Note also that in the strict Regge limit $p_3\to p_1$ and $p_4\to p_2$
and therefore one could express \cref{eq:regge2} in terms of two
Wilson lines spanning from $-\infty$ to $+\infty$ along the direction of the 
incoming particles.

Once again, the product of Wilson lines 
can be recast in term of a single exponential, which takes 
the form 
\cite{Melville:2013qca}
\begin{align}
{\cal A}_{\text{E}}&=\exp\left[K(z)\left(i\pi 
s+t\log\left(\frac{s}{-t}\right)\right)\right]
=e^{i\chi_{\text{E}}}\left(\frac{s}{-t}\right)^{K(z)t}~,
\label{regge}
\end{align}
where $K(z)$ is an infrared divergent constant that in dimensional 
regularization in $d=4-2\epsilon$ dimensions reads
\begin{align}
K(z)=-\left(\frac{\kappa}{2}\right)^2\frac{
	(\mu^2z^2)^{\epsilon}
}{{8\pi}^{2-\epsilon}}
\frac{
	\Gamma(1-\epsilon)
}{\epsilon}~.
\end{align}
The first term in the exponent of \cref{regge} is  
the so-called \emph{eikonal phase} $\chi_{\text{E}}$, while 
the second term instead is subleading in 
$t/s$ and 
contains information about the Regge trajectory of the graviton.

As the above discussion emphasized, both the soft 
and the eikonal exponentiation become particularly clear in the Wilson line 
description. In fact, the Wilson line 
generates soft emissions along the classical straight worldline of the hard 
particle
to all order in the coupling constant. Hence, it is by itself another kind of 
exponentiation, that we dubbed \emph{worldline} exponentiation.
In order to achieve a soft and eikonal exponentiations at subleading power, one 
can follow the same pattern: one first derives a worldline exponentiation with 
 GLWs, which are then combined at the amplitude level in a VEV 
using the full 
four-dimensional theory. 

Specifically, using a standard notation in the literature, one can generalize 
\cref{eq:softfunct} with a \emph{next-to-soft} function
${\cal \widetilde S}$ defined as
\begin{align}
{\cal \widetilde S}&=\langle 0 |\widetilde W_{p_1}(-\infty,0)
\widetilde W_{p_2}(-\infty,0)\widetilde W_{p_3}(0,\infty)\widetilde 
W_{p_4}(0,\infty)|0\rangle~,
\label{eq:softfunct2}
\end{align} 
where $\widetilde W_{p}(0,\infty)$ has been defined in \cref{gwl}. Similarly,
\cref{eq:regge2} is generalized with 
a \emph{next-to-eikonal} function 
\begin{align}
{\cal A}_{\text{NE}}&=\langle 0 
|
\widetilde W_{p_1}(0,-\infty,0)
\widetilde W_{p_2}(z,-\infty,0)
\widetilde W_{p_3}(0,0,\infty)
\widetilde W_{p_4}(z,0,\infty)
|0\rangle~,
\label{regge3}
\end{align} 
where the definition for $\widetilde W_{p}(z,0,\infty)$ follows from 
\cref{gwl} by shifting the argument of the graviton field in 
analogy with 
\cref{wilsonZ}. 
The exponentiation of ${\cal \widetilde S}$ and ${\cal A}_{\text{NE}}$ then
follows from the 
worldline exponentiation of the GWL.

\subsection{Classical limit}

At this point we note that unlike the standard amplitude program for classical 
gravitational scattering, we haven't considered any classical limit $ \hbar\to 
0$.
In fact, the GWL is not constructed via the classical limit but rather the soft 
one. However, there is growing evidence that the soft limit 
is intimately connected with the Regge limit, where the exchanged graviton is 
soft.
In particular, the classical information of the full quantum amplitude is 
contained in the eikonal phase, which, as we have mentioned, forms the leading 
contribution in the Regge limit of gravitational amplitudes.

\begin{figure}
	\centering
	\includegraphics[width=160mm]{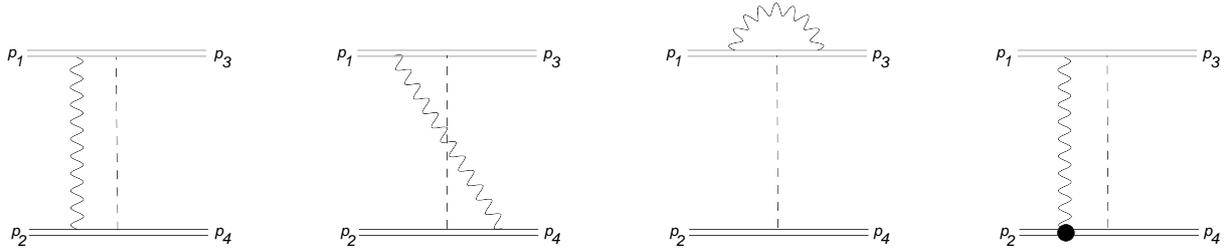}
	\caption{Selection of 1PM diagrams arising from the VEV of GWLs in 
		\cref{regge3}. All vertices are eikonal, while the blob represents a 
		next-to-eikonal vertex. 
		Only the leftmost diagram contributes to the eikonal phase. All 
		diagrams 
		contribute to the Regge trajectory of the graviton.
		The dashed line splits each Wilson line in two 
		branches corresponding to the 
		initial and final states of the 
		leading order amplitude. In position space, this affects the 
		limits of integration over 
		the position of the GWL vertices. }
	\label{fig:eik}
\end{figure}

Yet, computing the VEVs of GWLs is in general a quantum 
computation that contains information about both the 
eikonal phase and the Regge trajectory, as shown in \cref{regge}. 
Therefore, this approach does not seem to provide an efficient method for the 
selection of 
terms 
surviving 
the classical limit 
 at the integrand level, i.e. a method not relying on taking the limit 
 $\hbar\to 0$ of 
 the 
 full 
 quantum result. In fact, selecting only the diagrams that 
 contribute
 to the (next-to) eikonal phase seems challenging at higher orders in  
 Post-Minkowskian (PM) expansion.

\begin{figure}
 	\centering
 	\includegraphics[width=110mm]{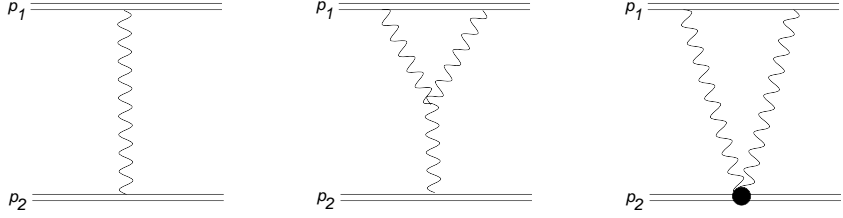}
 	\caption{Diagrams contributing to (next-to-)eikonal 
 		phase up to 2PM. Since 
 		the strict Regge limit $s\to\infty$ is considered, the hard particles 
 		do not recoil and the direction of each Wilson line
 		is specified by the initial momenta. 
 		Therefore, 
 		no dashed line is necessary and  
 		the integration over the position of the GWL vertices runs from 
 		$-\infty$ to $+\infty$. }
 	\label{fig:eik2}
 \end{figure}

 However, the description with (G)WLs offers the advantage of a diagrammatic 
 approach to tackle this problem. In order to illustrate this point, let us 
 consider 
%
%
   the first diagrams in \cref{fig:eik}, which in position space reads
   \begin{align}
   \kappa^2\,
   p_1^{\mu}\,p_1^{\nu}\,p_2^{\rho}\,p_2^{\sigma}
   \int_{-\infty}^0d\lambda_1\int_{-\infty}^0d\lambda_2\,
   P_{\mu\nu\rho\sigma}(\lambda_1 p_1-\lambda_2p_2-z)~,
   \label{eq:diag-a}
   \end{align}
   where $P_{\mu\nu\rho\sigma}(x)$ is the graviton propagator in position space.
   Following \cref{regge3}, here we have distinguished initial and final states 
   of 
   the 
   leading 
   order amplitude, as can be seen by the upper integration limits being 
   zero, rather than $+\infty$. Diagrammatically, this can be represented with 
   a dashed line splitting each Wilson line into two branches.  
    Obviously, one has to add to \cref{eq:diag-a} the other three diagrams 
   corresponding 
   to the remaining integration regions in $\lambda_1$ and $\lambda_2$. 
   The sum of these four diagrams is equivalent to a single diagram where both 
   integration limits 
   span from $-\infty$ to $+\infty$ and no dashed line is necessary, as 
   depicted in the first diagram in \cref{fig:eik2}.
   However, by considering only diagrams without a dashed line, we would miss 
   the 
   third diagram in \cref{fig:eik}, 
   which is necessary in order to generate logarithms of $s/t$, rather than 
   $s/m^2$ \cite{Melville:2013qca}.
        
   However, if we are interested only in the classical limit, one can consider 
   the strict 
   $s\to \infty$ limit and represent the diagrams without distinction between 
   the initial and final states of the leading order amplitude (i.e. by setting 
   $p_3\to p_1$ and $p_4\to p_2$).
 In fact, it has 
 been known for a 
 long time that at leading order in the soft 
 expansion the 
 eikonal phase can be extracted from diagrams where the graviton connects 
 either incoming or outgoing particles (as the first diagram in 
 \cref{fig:eik}).  
 On the other hand, for the Regge trajectory one has to include also diagrams 
 where 
 initial and final legs are connected (as the second and third diagrams in 
 \cref{fig:eik})
  \cite{Kabat:1992tb}.
 This property is in agreement with the well-known fact that the eikonal phase 
 arises from loop corrections to the cusp angle in 
 spacelike 
 kinematics \cite{Laenen:2015jia, Magnea:1990zb}. In fact, one can observe that 
 the impact 
 parameter $z$ is a UV 
 regulator 
 for the amplitude in \cref{eq:regge2}. Thus, by setting it to zero one 
 recovers 
 the four-point soft function of \cref{eq:softfunct} in the forward limit, 
 where gravitons connecting either initial or final legs correspond to 
 loop corrections in the spacelike kinematics.

 Therefore, at leading order in the soft expansion the eikonal phase is given 
 by the first diagram in \cref{fig:eik2} in the Regge limit. 
 The fact that this diagram matches 
 the leading order amplitude, i.e. ${\cal A}_{\text{LO}}=2is\chi_{\text{E}}$, 
 should come of 
 no surprise, given that the exchanged graviton in ${\cal A}_{\text{LO}}$ is 
 also 
 soft. The merit of the derivation with Wilson lines is that the 
 exponentiation of this diagram becomes manifest,
 since one can write
 \begin{align}
 e^{i\chi_{\text{E}}}=\langle 0 
 |W_{p_1}(0,-\infty,\infty)
 W_{p_2}(z,-\infty,\infty)
 |0\rangle ~,
 \label{eq:regge2bis}
 \end{align}
  where we made use of the fact that in the strict Regge limit \cref{eq:regge2}
 can be written in terms of two Wilson lines spanning from
 $-\infty$ to 
 $+\infty$.


  Things 
 are more difficult at subleading power in the eikonal 
 expansion \cite{Akhoury:2013yua}. Indeed, while the seagull vertex
 in \cref{fig:eik2} 
 is of order $\kappa^2$ and thus corresponds 
 to a Post-Minkowskian correction (i.e. $G/z$) to the eikonal phase, 
 single emissions like the one of \cref{NEvertex} (e.g. the third diagram in 
 \cref{fig:eik})) bring 
 quantum information as $G/z$ corrections to the Regge trajectory. The main 
 problem is that  
 only after all integrals have been performed it becomes clear
 what contributes to the eikonal phase or the Regge trajectory 
 \cite{Luna:2016idw}. This makes the extraction of the classical 
 limit less  
 efficient.

 On the other hand, the derivation of the GWL in this paper is based on the 
 Hamiltonian of \cref{xp} (and subsequently on the Lagrangians in  
 \cref{asympt-exact} and \cref{asympt-expanded}). Restoring the explicit 
 dependence on $\hbar$ in \cref{xp}, one can appreciate that
 the terms of order $V^{\mu}$ defined in \cref{counter}
 are subleading in $\hbar$. In fact, the Lagrangian of \cref{asympt-exact}
 becomes
 \begin{align}
 L[\tilde x,a,b,c]&=- 
 \frac{1}{2}
 \left(
 (\dot{\tilde x}^{\mu}\dot {\tilde x}^{\nu}
 +  {a}^{\mu}  {a}^{\nu}
 +  {b}^{\mu}  {c}^{\nu})
 g_{\mu\nu}
 + i\hbar  (\dot {\tilde x}+p_f)^{\mu}g_{\mu\nu}V^{\nu}
 -\frac{\hbar^2}{4}V^{\mu}g_{\mu\nu}V^{\nu}
 \right) \notag \\
 &=- 
 \frac{1}{2}
 (\dot{\tilde x}^{\mu}\dot {\tilde x}^{\nu}
 +  {a}^{\mu}  {a}^{\nu}
 +  {b}^{\mu}  {c}^{\nu})
 g_{\mu\nu}
 +{\cal O}(\hbar)~.
 \label{asympt-exact-hbar}
 \end{align}
After performing the weak field expansion, \cref{asympt-exact-hbar} implies 
that in the classical limit and at the next-to-soft level one can neglect 
the following term in \cref{asympt-expanded}:
\begin{align}
	-i\frac{\kappa}{2\lambda}
p_f^{\mu}
\left(
\frac{1}{2}\partial_{\mu}h-\partial_{\nu}h_{\mu\nu}
\right)~.
\label{quantum}
\end{align}
Moreover, one should neglect quantum fluctuations along the worldline, and 
therefore set to zero the equal-time propagators. The net effect is that all 
terms in \cref{NEpath} do not contribute in the classical limit. 
As explained in detail in \cref{sec:exp}, these give rise to 
the single
NE emission of  
\cref{NEvertex}. The corresponding diagrams at the amplitude level 
in the 
Regge limit (e.g. the diagram on the right 
in \cref{fig:eik}) contribute to 
the Regge 
trajectory but not to the eikonal phase \cite{Luna:2016idw}, hence it is a pure 
quantum effect.
The Lagrangian in \cref{asympt-exact-hbar} makes it clear why these 
contributions can be discarded in the classical limit at the integrand level.  
Both the single eikonal vertex in \cref{Evertex} and the seagull NE vertex of 
\cref{NEvertices}, on the other hand, originate from 
terms in \cref{asympt-exact-hbar} that are leading in the $\hbar$ expansion and 
indeed they contribute to the eikonal phase, as shown in \cref{fig:eik2}.

In conclusion, the exponentiated next-to-eikonal phase $\chi_{\text{NE}}$ (i.e. 
the 
eikonal phase modified by subleading power corrections) can be 
computed in the strict Regge limit via
\begin{align}
e^{i\chi_{\text{NE}}}&=
\langle 0 
|\widetilde W_{p_1}^{\text{cl}}(0,-\infty,\infty)\widetilde 
W_{p_2}^{\text{cl}}(z,-\infty,\infty)|0\rangle~,
\label{eq:regge3}
\end{align}
where we isolated the classical contribution in the GWL of \cref{gwl} by 
defining 
\begin{align}
\widetilde W_p^{\text{cl}}(z,-\infty,\infty)&\equiv\exp\Bigg\{
-\frac{i\kappa}{2}\int_{-\infty}^{\infty} dt\, p_{\mu}p_{\nu}
h^{\mu\nu}(pt+z)
\notag \\
& + \frac{i\kappa^2}{2}\int_{-\infty}^{\infty} dt \int_{-\infty}^{\infty} 
ds\,
\Bigg[
\frac{p^{\mu}p^{\nu}p^{\rho}p^{\sigma}}{4}\min(t,s)\,\partial_{\alpha}
h_{\mu\nu}(pt+z)
\partial^{\alpha}
h_{\rho\sigma}(ps+z)
\notag\\&\qquad
+p^{\mu}p^{\nu}p^{\rho}\,\theta(t-s)\,
h_{\rho\sigma}(ps+z)
\partial_{\sigma}
h_{\mu\nu}(pt+z)
\notag\\&\qquad 
\,+\,p^{\nu}p^{\sigma}\,\delta(t-s)\,
h^{\mu}_{\;\,\sigma}(ps+z)
h_{\mu\nu}(pt+z)
\Bigg]\Bigg\}~.
\label{gwl-cl}
\end{align} 
Finally, note that although we have discussed the eikonal phase only up 
to 2PM, 
this 
way of extracting the classical limit has the 
potential of extending it to 
higher orders in the PM expansion.

 \subsection{A brief comparison with the WQFT approach}
 \label{sec:comp}
 
 It is instructive to compare the GWL approach with a recent body of work 
 \cite{Mogull:2020sak, Jakobsen:2021smu, Jakobsen:2021lvp, Jakobsen:2021zvh} 
 investigating classical scattering with worldline techniques, hence dubbed 
 Worldline Quantum Field Theory (WQFT).
 Considering the scalar case and the original work of White 
 \cite{White:2011yy}  
 which we have revisited in \cref{sec:review}, one could 
 conclude that a representation without ghosts fields and based on an intricate 
 Hamiltonian defined only in the weak field expansion is rather different to 
 the WQFT description of \cite{Mogull:2020sak}. 
 However, the two approaches are equivalent 
 in the classical limit as can be seen from the derivation  
 in \cref{sec:worldline}. 
 Let us 
 discuss this point in greater detail. 
  
 The set up is similar. The authors of \cite{Mogull:2020sak}
 consider a complex scalar field non-minimally coupled to gravity via
 \begin{align}
 S=\int d^dx\, \sqrt{-g}\left(
 g^{\mu\nu}\partial_{\mu}\phi^* \partial_{\nu}\phi 
 -m^2|\phi|^2
 +\xi R |\phi|^2
 \right)
 ~,
 \label{xi}
 \end{align}
where $R$ is the scalar curvature. 
A worldline representation for the dressed propagator is then built in 
configuration space. Choosing  path 
integrals in configuration space as a starting point hides the quantization 
procedure in phase 
space and the corresponding Hamiltonian, that we have discuss in detail in 
\cref{sec:worldline}. 
This information can be 
extracted from the counterterm $R(x)/4$ that the authors of 
\cite{Mogull:2020sak}  add by hand to the 
interacting Lagrangian. This is the counterterm that naturally arises
in covariant regularization schemes with the Hamiltonian of \cref{Hherm} in 
Weyl-ordered form 
\cite{Bastianelli:2006rx}. Therefore, unlike the time slicing quantization in 
$px$-ordering  
discussed 
in this work (and in \cite{White:2011yy}), the quantization 
procedure 
underlying the construction of 
the WQFT is the one that has been discussed in detail in the 
literature. Moreover, the calculations explicitly presented in 
\cite{Mogull:2020sak} are for $\xi=1/4$ and in de-Donder gauge, so that the 
worldline action in configuration space for the $x$ field assumes the trivial 
form 
\begin{align}
-\frac{1}{4}\int dt 
\left({\dot x}^{\mu}{\dot x}^{\mu}
+a^{\mu}a^{\nu} +b^{\mu}c^{\nu}\right)g_{\mu\nu}~,
\end{align}
 which can be compared with the gauge invariant 
expression  in \cref{asympt-exact} in this work (which considers $\xi=0$). 
However, as \cref{asympt-exact-hbar} 
and 
the related discussion make clear, all these differences disappears in the 
classical limit, since all counterterms are suppressed in the $\hbar$ 
expansion. Therefore, the GWL and the WQFT are completely equivalent at the 
Lagrangian level in the classical limit.

A minor difference emerges in the boundary conditions of the path integral. 
Indeed,
 the amplitude in the WQFT formalism is constructed with two dressed 
 propagators extending from 
 $-\infty$ to $+\infty$, in the same spirit as the high energy description in 
 terms Wilson lines of \cref{eq:regge2bis}. This is the reason for a symmetric 
 representation that accommodates well Weyl-ordered Hamiltonians. The GWL, on 
 the other hand, is constructed from a dressed propagator emerging from a 
 localized hard interaction. This is to make it suitable to describe not only 
 high energy scattering (and the corresponding classical limit) but also the 
 soft exponentiation of \cref{eq:softfunct} and the subleading Regge trajectory 
 in \cref{regge}. However, as \cref{eq:regge2bis}
 and \cref{gwl-cl} make
 clear, the representation of the GWL can be used also for lines extending from 
 $-\infty$ to $+\infty$. 

The main difference between the two approaches  
is evident only when moving to the full scattering amplitude. 
In fact, in the approach used in this paper  
the 
worldine path integral is solved once  for each GWL. The GWLs are then 
combined in a VEV like \cref{eq:regge2} using the properly gauge-fixed 
Einstein-Hilbert action,  
i.e. in a path 
integral over the soft graviton field. 
The WQFT, on the other hand, is defined with a mixed worldline and 
four-dimensional description of the scattering amplitude. More specifically, 
by leaving both the path integral over the graviton and the worldline fields 
unsolved, a master formula is derived for the full amplitude.
This has the advantage that one can express classical 
observables such as the impulse or the radiated momentum in terms of worldline 
fields. 
However, for practical purposes, we note that this is a minor difference, 
since the integrals that one has to solve are the same (i.e. the diagrams in 
\cref{fig:eik2}).
More specifically, in both cases the classical limit is extracted by a 
diagrammatic analysis rather than a 
power counting in $\hbar$.

Finally, we note that both in the GWL and in the WQFT approaches one expands 
the worldline path integral
around the classical straight path. Given the well-known role of Wilson lines 
in describing soft emissions, the GWL representation emphasizes that this 
corresponds to a soft  expansion of the exchanged or radiated graviton, but it 
is clear that from a 
computational point of view this is exactly the same. We note however that 
factorization breaking terms given by Low's theorem are not immediately clear 
without recurring to a soft argument. As remarked in \cref{sec:fact}, these 
correspond to soft emissions sensitive to the angular momentum of the scalar 
particle, which in turn can be described in terms of the Hard (i.e. Leading 
Order) dynamics with shifted kinematics \cite{Akhoury:2013yua}. However, the 
corresponding contribution to the eikonal phase vanishes at 2PM 
\cite{Luna:2016idw} and therefore are irrelevant for the computations decribed 
in \cite{Mogull:2020sak}.

\section{Conclusions}
The  Generalized Wilson Line (GWL) is a powerful tool to describe all-order 
properties of scattering amplitudes at subleading power in the 
soft expansion. This operator was originally introduced in 
 gauge theories by Laenen, Stavenga and White \cite{Laenen:2008gt} and in 
 gravity by White \cite{White:2011yy}. Later it was applied in 
 the computation of the high energy limit of scattering amplitudes
 at the 
  next-to-soft level \cite{Luna:2016idw}.  
By rederiving the GWL 
 from a worldline model in curved spacetime, in this paper we have clarified a 
 number of issues that were not addressed in the literature. 
 
Firstly, we have discussed in great detail the regularization of UV 
 divergences. Although these techniques have been known for a long time, we 
 have applied them for the first time in the unusual setting required by the 
 GWL, i.e. in the asymptotic limit on a infinite worldline in a mixed 
 position-momentum representation. In doing so, we
 have  unambiguously identified the Hamiltonian in 
 $px$-ordering (i.e. \cref{xp}) underlying the construction of the GWL. We have 
 also
  carefully derived the worldline correlators in time slicing 
 regularization.   At the price of setting up a 
 quantization with a 
 non-hermitian conjugate momentum, 
 this procedure leads to a much simpler expression for the worldline Lagrangian 
 (\cref{asympt-expanded}) compared with the equivalent expression 
  in the 
 earlier derivation (\cref{action}).
 
 Secondly, we have discussed the role of the GWL in the extraction of the 
 classical 
 limit of scattering amplitudes. Indeed, one of the virtues of the worldline 
 model in curved spacetime is 
 that one can easily isolate terms that are suppressed in $\hbar$ at the 
 Lagrangian level. In this way we have identified the terms (\cref{NEvertex}) 
 that can be discarded 
 in the classical limit. This is in agreement with ref. \cite{Luna:2016idw},  
 where the 
 corresponding contribution at the amplitude level was found to be relevant for 
 the (quantum) Reggeization but not for the classical eikonal phase.
 
Finally, we have compared the GWL approach with the recently proposed 
Worldline 
Quantum Field Theory (WQFT) \cite{Mogull:2020sak}. The 
derivation of the GWL  from first-principles achieved in this work shows 
explicitly that the two approaches 
are equivalent in the classical limit. In particular, they give rise to the 
same class of diagrams at the amplitude level, although in the GWL approach the 
worldline path integral is solved separately for each external leg. We stress 
however that the GWL is also suitable for the investigation of purely quantum 
properties 
such as the Reggeizaition of the graviton.    

There are a number of directions in which this work can be extended. The most 
obvious one is the derivation of the GWL for spinning emitters, which is a 
topic of 
high priority given the great demand for precision 
calculations in gravitational 
physics \cite{Levi:2015ixa, Levi:2018nxp, Levi:2020lfn, 
Vines:2016qwa, 
Guevara:2018wpp, Guevara:2020xjx}. As remarked in \cref{sec:intro}, the 
worldline model in 
curved 
space discussed in this paper provides a natural set up for such a 
generalization.  Work in this direction is ongoing 
 \cite{spinpaper}. Another interesting direction is a detailed study of real 
 radiation effects, which can be easily accommodated in the GWL approach by 
 inserting graviton fields in the VEV of \cref{eq:softfunct2} and \cref{regge3}.
  More generally, the GWL provides a tool that might be 
 useful to investigate other aspects of high energy scattering, such as the 
role of the classical double copy \cite{Monteiro:2014cda} at higher powers in 
the soft expansion.

\section*{Acknowledgments}
The authors would like to thank Olindo Corradini, Gustav Mogull, Jan Plefka, 
Jan Steinhoff and 
Chris White for 
useful discussions and correspondence.

\appendix

\section{Worldline diagrams for the GWL in \cref{sec:review}}
\label{sec:diag}
In this section we compute explicitly the connected diagrams arising from the 
evaluation of  
\cref{finfty} at the next-to-soft level. 

At leading order in $1/\lambda$ one considers only the eikonal term 
\begin{equation}\label{}
-\frac{i\kappa}{2}\int\limits_{0}^{\infty}\lambda h_{\mu\nu}p^\mu 
p^\nu=-\int\limits\frac{d^4k}{(2\pi)^4}\tilde{h}_{\mu\nu}\frac{\kappa}{2}\frac{p^\mu
 p^\nu}{pk}~,
\end{equation}
in agreement with \cref{Evertex}.

At order $\lambda^0$ we start including quantum fluctuations in \cref{finfty}.
These arise by 
expanding the graviton field in \cref{action} as
\begin{align}
h_{\mu\nu}(pt+x)=h_{\mu\nu}(pt)
+x^{\rho}\partial_{\rho}h_{\mu\nu}(pt)
+\frac{1}{2}x^{\rho}x^{\sigma}\partial_{\rho}\partial_{\sigma}h_{\mu\nu}(pt)
+{\cal O}(x^3)
\end{align}
and subsequently 
connecting vertices with powers of $x$ with
the correlators in \cref{corr1}, \cref{corr2} and \cref{corr3}.
We distinguish one-graviton and two-graviton contributions.

For the former
we get
\begin{align}
&\nonumber\frac{i\kappa}{2}\int\limits_{0}^{\infty}dt~\Bigg(-\frac{\lambda}{2} 
\partial_\rho\partial_\sigma h_{\mu\nu}p^\mu p^\nu \wick{\c x^\rho(t)\c 
x^\sigma(t)}+i(\partial^\mu h_{\mu\nu}p^\nu-\frac{1}{2}\partial_\mu 
hp^\mu)\Bigg)\\
=&\int\limits\frac{d^4k}{(2\pi)^4}\tilde{h}_{\mu\nu}(k)~\left[-\frac{\kappa}{4}\frac{k^2p^\mu
 p^\nu}{(pk)^2}+\frac{\kappa}{4}\frac{p^\mu k^\nu+p^\nu 
k^\mu-pk\eta^{\mu\nu}}{pk}\right]~,
\end{align}
in agrement with \cref{NEvertex}.

The calculations of the two-graviton contribution  involves four terms.  
The first one does not involve any correlator of $x$ and reads 
\begin{align}
&\nonumber 
i\frac{\kappa^2}{4}\lambda\int\limits_{0}^{\infty}dt\left(hh_{\mu\nu}p^\mu 
p^\nu-\frac{h^2}{2}m^2\right)\\
=&\int\limits \frac{d^4kd^4l}{(2\pi)^8} 
\frac{\tilde{h}_{\mu\nu}(k)\tilde{h}_{\rho\sigma}(l)}{2} 
\left[\frac{\kappa^2}{4p(k+l)}\left(2\eta^{\rho\sigma}p^\mu 
p^\nu-m^2\eta^{\mu\nu}\eta^{\rho\sigma}\right)\right]~.
\label{one}
\end{align}
The second term requires the correlator $\langle xx\rangle$ and gives
\begin{align}
&\nonumber\frac{i^2}{2}\int\limits_{0}^{\infty}dtdt' 
\frac{\kappa^2}{4}\lambda^2~\left(\partial_\alpha h_{\mu\nu}\wick{\c 
x^\alpha(t) p^\mu p^\nu~~\partial_\beta h_{\rho\sigma}\c 
x^\beta(t')}p^\rho p^\sigma\right)\\
=&\nonumber\int\limits \frac{d^4kd^4l}{(2\pi)^8} 
\frac{\tilde{h}_{\mu\nu}(k)\tilde{h}_{\rho\sigma}(l)}{2}\left[\frac{-i\kappa^2}{4}kl~p^\mu
 p^\nu p^\rho p^\sigma\right]\int\limits_{0}^{\infty}dtdt' 
\min(t,t')e^{-ikpt}e^{-ilpt'}\\
=&\int\limits \frac{d^4kd^4l}{(2\pi)^8} 
\frac{\tilde{h}_{\mu\nu}(k)\tilde{h}_{\rho\sigma}(l)}{2}\left[\frac{\kappa^2}{4}\frac{kl}{pl~pk~p(k+l)}~p^\mu
 p^\nu p^\rho p^\sigma\right].
 \label{two}
\end{align}
The third term requires the correlator $\langle \dot x\dot x\rangle$ and gives
\begin{align} 
&\nonumber\frac{i^2}{2}\int\limits_{0}^{\infty}dtdt' 
\frac{\kappa^2}{4}\lambda^2~\left(-2h_{\alpha\nu}p^\nu+hp_\alpha\right)
\left(-2h_{\beta\nu}p^\nu+hp_\beta\right)\wick{\dot{x}^\alpha\c(t)
 \dot{x}^\beta\c(t')}\\
=&\nonumber\int\limits \frac{d^4kd^4l}{(2\pi)^8} 
\frac{\tilde{h}_{\mu\nu}(k)\tilde{h}_{\rho\sigma}(l)}{2}
\left[\frac{i\kappa^2}{4}\eta_{\alpha\beta}\left(-2\eta^{\mu\alpha}p^\nu\!+\!\eta^{\mu\nu}
 p^\alpha\right)\!\left(-2\eta^{\rho\beta}p^\sigma\!+\!\eta^{\rho\sigma} 
p^\beta\right)\right]\notag\\
&\quad \int\limits\displaylimits_{0}^{\infty}dtdt' 
\delta(t\!-\!t')e^{-ikpt}e^{-ilpt'}\notag\\
=&\int\limits \frac{d^4kd^4l}{(2\pi)^8} 
\frac{\tilde{h}_{\mu\nu}(k)\tilde{h}_{\rho\sigma}(l)}{2}\left[\frac{\kappa^2}{4~p(k+l)}\left(4\eta^{\mu\rho}p^\nu
 p^\sigma+m^2\eta^{\mu\nu}\eta^{\rho\sigma}-2(\eta^{\mu\nu}p^\rho 
p^\sigma+\eta^{\rho\sigma}p^\mu p^\nu)\right)\right]~.
\label{three}
\end{align}
The fourth term requires the correlator $\langle x\dot x\rangle$  and gives
\begin{align}
&\nonumber i^2\int\limits_{0}^{\infty}dtdt' 
\frac{\kappa^2}{4}\lambda^2~\left(-\partial_\alpha h_{\mu\nu}\wick{\c 
x^\alpha(t) p^\mu p^\nu\left(-2h_{\beta\nu}p^\nu+hp_\beta\right) 
\dot{x}^\beta\c(t')}\right)\\
=&\nonumber\int\limits \frac{d^4kd^4l}{(2\pi)^8} 
\frac{\tilde{h}_{\mu\nu}(k)\tilde{h}_{\rho\sigma}(l)}{2}\left[-\frac{\kappa^2}{2}\left(-2k^\rho
 p^\sigma p^\mu p^\nu+kp\eta^{\rho\sigma}p^\mu 
p^\nu\right)\right]\int\limits_{0}^{\infty}dtdt' 
\theta(t-t')e^{-ikpt}e^{-ilpt'}\\
=&\int\limits \frac{d^4kd^4l}{(2\pi)^8} 
\frac{\tilde{h}_{\mu\nu}(k)\tilde{h}_{\rho\sigma}(l)}{2}\left[\frac{\kappa^2}{2~kp~(k+l)p}\left(-2k^\rho
 p^\sigma p^\mu p^\nu+kp\eta^{\rho\sigma}p^\mu p^\nu\right)\right]~.
\end{align}
After combining the four terms, the integrand reads
\begin{equation}
\frac{\kappa^2}{4}\frac{1}{p(k+l)}\left[2\eta^{\mu\nu}p^\rho p^\sigma 
-2\eta^{\rho\sigma}p^\mu p^\nu+\frac{kl}{pl~pk}p^\mu p^\nu p^\rho 
p^\sigma-\frac{4k^\rho p^\sigma p^\mu p^\nu}{kp}+4p_\mu 
p_\rho\eta_{\nu\sigma}\right]~.
\label{finalrmunu}
\end{equation}
Upon symmetrizing the expression with respect to 
$(k;\mu\nu)\leftrightarrow(l;\rho\sigma)$ and 
$\mu\leftrightarrow\nu~,~\rho\leftrightarrow\sigma$, the first two terms 
in \cref{finalrmunu} cancel and we are left precisely with the contribution of
\cref{NEvertices}.

\section{Derivation of worldline correlators}
\label{sec:corr}

The derivation of the two-point correlators of the fields $x^\mu(t)$, 
$a^\mu(t)$, $b^\mu(t)$ and $c^\mu(t)$ can be obtained straightforwardly in the 
continuum, 
given the 
simple nature of the kinetic terms. However, as remarked in 
\cref{sec:worldline}, these Green functions become ambiguous at equal time. 
No ambiguity is left if the calculation is carried in the discretized version, 
as we discuss in this appendix. Given that we work in $px$-ordering and that we 
consider the asymptotic limit $T\to\infty$, the derivation is slightly 
different 
from similar calculations available in the literature.
In doing so, we will use some trigonometric identity whose proof is outlined in 
\cref{sec:trig}. 
We compute the two-point correlators for the ghosts and $x$ fields separately.

\subsection{Ghost correlators}
\label{sec:ghost}
The ghost sector does not present any particular difficulty. 
We consider the time sliced path integral representation of 
\cref{Zghosts} after performing the weak field expansion. 
Introducing a source term for each 
ghost field we get 
\begin{align}
Z_{ghosts}[A,B,C]\equiv& \int\limits \prod_{i=1}^{N-1}d^4a_id^4b_id^4c_i  
\exp\Bigg\{-\frac{\tau}{2}\sum_{n=1}^{N-1}\eta_{\mu\nu}^n(a_n^\mu 
a_n^\nu+b^\mu_nc^\nu_n)\\
&
-\frac{\tau}{2}\sum_{n=1}^{N-1}\kappa\,h_{\mu\nu}^n(a_n^\mu 
a_n^\nu+b^\mu_nc^\nu_n)
+\sum_{i=1}^{N-1}\left(A_\mu^ia_i^\mu+b_i^\mu 
B^i_\mu+C_\mu^ic^\mu_i\right)\Bigg\}~.
\end{align}
After factorizing the free part $Z_{0}^{ghosts}[A,B,C]$
from the interacting part $Z_{int}^{ghosts}$, and performing the Gaussian 
integration in the former, we get   
\begin{align}
Z_{ghosts}[A,B,C]&=\exp\left\{\frac{\tau}{2}
\sum_{n=1}^{N-1}\kappa\,h_{\mu\nu}(x_i)
\left(-\frac{\partial}{A_\mu^i}\frac{\partial}{A_\nu^i}
+\frac{\partial}{B_\mu^i}\frac{\partial}{C_\nu^i}\right)\right\} 
\exp\left\{\frac{\eta_{\mu\nu}}{\tau}
\sum_{i=1}^{N-1}\frac{1}{2}A_\mu^iA_\nu^i+2C_\nu^iB_\mu^i\right\} \notag \\
&=Z_{int}^{ghosts}Z_{0}^{ghosts}[A,B,C],
\label{ghostZ}	
\end{align}
where we assumed the left differentiation convention when differentiating 
w.r.t. our Grassmann variables. 
All correlators can be easily calculated from \cref{ghostZ}. In particular,  
the propagators are given by 
\begin{align}\label{}
\wick{\c a_i^\mu \c a_j^\nu}&=\frac{\partial}{\partial 
A_\mu^i}\frac{\partial}{\partial 
A_\nu^j}Z_{0}^{ghosts}[A,B,C]\Big\rvert_{A=B=C=0}=\frac{1}{\tau}\eta^{\mu\nu}\delta_{ij}~,\\
\wick{\c b_i^\mu \c c_j^\nu}&=-\frac{\partial_L}{\partial 
B_\mu^i}\frac{\partial_L}{\partial 
C_\nu^j}Z_{0}^{ghosts}[A,B,C]\Big\rvert_{A=B=C=0}=-\frac{2}{\tau}\eta^{\mu\nu}\delta_{ij}~.
\end{align}
This tells us that in the continuum limit they are given by 
\begin{align}\label{}
\wick{\c a(t)^\mu \c a(t')^\nu}&=\eta^{\mu\nu}\delta(t-t')~,\\
\wick{\c b(t)^\mu \c c(t')^\nu}&=-2\eta^{\mu\nu}\delta(t-t')~,
\end{align}
in agreement with \cref{ghost1}, \cref{ghost2}, \cref{corra} and \cref{corrbc}  
in the main text.

\subsection{Correlators for $x(t)$}
\label{sec:x}
Things are more difficult for the $x$-correlators.
We consider the relevant part in
\cref{Zdiscr}, which reads
\begin{align}\label{Zdiscr2}
e^{ip_Nx_N}\int\limits \prod_{i=2}^{N}\frac{d^4x_i}{(2\pi\tau)^n} 
\exp\left\{\frac{i}{2\tau}\sum_{n=1}^{N-1}g_{\mu\nu}^n
B^\mu_n B^\nu_n\right\}~,
\end{align} 
where 
\begin{align}
B^\mu_n&\equiv -i\tau 
\left[\frac{(x_{n+1}-x_n)^\mu}{\tau}
+\frac{i}{2}\Big(\partial_{\nu}g^{\mu\nu}(x_n)+g^{\mu\nu}(x_n)
(\partial_{\nu}\ln(\sqrt{-g(x_n)}))\Big)
\right]~.
\end{align}
The strategy to work out the propagators is the same as the ghost 
case, i.e. introducing a source $F_{\mu}^i$ for each field and functional 
differentiating 
after solving the path integral. However, this time the action is not diagonal 
w.r.t. the index $i$. 
In fact, 
 isolating  the term quadratic in $\tilde{x}$  in \cref{Zdiscr2}  
and adding a 
source term for the quantum fluctuation yields
\begin{equation}\label{discr-exp}
\int\limits \prod_{i=1}^{N-1} \frac{d^4x_i}{(2\pi\tau)^2} 
\exp\left\{-\frac{i}{2\tau}\sum_{i=1}^{N-1}\eta_{\mu\nu}
(\tilde{x}_{i}-\tilde{x}_{i-1})^\mu
(\tilde{x}_{i}-\tilde{x}_{i-1})^\nu+\sum_{i=1}^{N-1}
F_\mu^i\tilde{x}^\mu_i\right\}~,
\end{equation}
where we have first re-indexed the sum in the exponential to match the 
product-measure, and then 
renamed the variables by shifting each index down by one 
(i.e. $x_N\rightarrow x_{N-1}$, ... , $x_1\rightarrow x_0$).

To perform the Gaussian integration we need to diagonalize the action in the 
index $i$. Following \cite{Bastianelli:2006rx}, we decompose the fields in 
modes  and write 
\begin{equation}\label{}
\tilde{x_i}=\sum_{m=1}^{N-1}q_m O_{mi}=\sum_{m=1}^{N-1}q_m 
\sqrt{\frac{2}{N}}\sin\left(\frac{im\pi}{N}\right)~.
\end{equation}
The matrix $O_{mi}$ is orthogonal, which means that 
$d^4x_i=d^4q_i$.
Moreover,
\begin{align}\label{}
O_{1i}&=O_{Ni}=0~,\\
O_{n,i+1}+O_{n,i-1}&=2O_{ni}\cos(\frac{n\pi}{N})~.
\end{align}
Therefore, the argument of the exponential in \cref{discr-exp} becomes
\begin{align}
&-\frac{i}{2\tau}\sum_{i=1}^{N-1}\eta_{\mu\nu}(\tilde{x}_{i}
-\tilde{x}_{i-1})^\mu(\tilde{x}_{i}-\tilde{x}_{i-1})^\nu
+\sum_{i=1}^{N-1}F_\mu^i\tilde{x}^\mu_i \notag \\
	&\quad =-\frac{i}{2\tau}\sum_{i,n,m=1}^{N-1}\eta_{\mu\nu}q^\mu_mq^\nu_n 
\left(2\delta_{mn}-O_{mi}o_{n,i+1}-O_{ni}O_{m,i+1}\right)
+\sum_{i,m=1}^{N-1}F_\mu^iO_{mi}q^\mu_m \notag \\
&\quad=\sum_{m=1}^{N-1}-\frac{1}{2}q_m^\mu q_m^\nu 
\frac{2i}{\tau}\eta_{\mu\nu}\left(1-\cos\left(\frac{m\pi}{N}\right)\right)
+\sum_{i,m=1}^{N-1}F_\mu^iO_{mi}q^\mu_m~.
\end{align}
Then, we can perform the Gaussian integration, to get
\begin{align}
&\prod_{m=1}^{N-1}\sqrt{\frac{(2\pi)^4}{(2\pi\tau)^4 
\det(\frac{2i}{\tau}(1-\cos(\frac{m\pi}{N}))\eta_{\mu\nu})}}
\exp\Bigg\{-\frac{i\tau}{4}\eta^{\mu\nu}\sum_{i,j,m=1}^{N-1}
\frac{F_\mu^iO_{im}O_{mj}F_\nu^j}{1-\cos(\frac{m\pi}{N})}\Bigg\} \notag \\
=&\prod_{m=1}^{N-1}\frac{1}{2^{2(N-1)}\sqrt{-1}(1-\cos(\frac{m\pi}{N}))^2}
\exp\Bigg\{-\frac{i\tau}{4}
\eta^{\mu\nu}\sum_{i,j,m=1}^{N-1}\frac{F_\mu^iO_{im}O_{mj}F_\nu^j}{1-\cos(\frac{m\pi}{N})}\Bigg\}.
\end{align}
Finally,  choosing $N-1=4m$ and making use of
\cref{b1}, we get
\begin{align}
&\frac{1}{N^2}\exp\Bigg\{-\frac{i\tau}{4}\eta^{\mu\nu}
\sum_{i,j,m=1}^{N-1}\frac{F_\mu^iO_{im}O_{mj}F_\nu^j}{1-\cos(\frac{m\pi}{N})}\Bigg\}
\equiv\frac{1}{N^2}Z_0\left[F\right]~.
\end{align}

We are now ready to compute the correlators. We start with 
the simplest one, i.e.
\begin{align}
\wick{\c x_k^\mu \c x_l^\nu}=&\frac{\partial}{\partial 
F_\mu^k}\frac{\partial}{\partial F_\nu^l}Z_0[F]\bigg\rvert_{F=0}\\
=&-\frac{i\tau}{2}\eta^{\mu\nu}\sum_{m=1}^{N-1}
\frac{O_{km}O_{ml}}{1-\cos\left(\frac{m\pi}{N}\right)}\\
=&-\frac{i\tau}{N}\left[-\frac{1}{2}
\sum_{m=1}^{l+k-1}\min(2k,2l,m)(1+(-1)^{l+k-m})+\frac{(N-1)}{2}\min(2k,2l)\right]\\
=&\frac{i\tau}{2N}\left[\sum_{m=1}^{l+k-1}\min(2k,2l,m)(1+(-1)^{l+k-m})-(N-1)\min(2k,2l)\right]
\end{align}
where in the third line we used  \cref{sinquotient}.

We can now evaluate the sum quite easily. Setting $k>l$ without loss of 
generality we get to
\begin{align}\label{}
&\sum_{m=1}^{l+k-1}\min(2k,2l,m)(1+(-1)^{l+k-m})-(N-1)\min(2k,2l)\\
=&\sum_{m=1}^{l+k-1}\min(2l,m)(1+(-1)^{l+k-m})-2l(N-1)\\
=&\sum_{m=1}^{2l-1}m(1+(-1)^{k+l-m})+2l\sum_{m=2l}^{k+l-m}(1+(-1)^{k+l-1})-2l(N-1)\\
\overset{k+l\text{ even}}{=}&\sum_{m\text{ even}\in 
\{1,...,2l-1\}}2m+4l\sum_{m\text{ even}\in\{2l,...,k+l-1\}}1-2l(N-1)=2l(k-N).
\end{align}
We obtain the exact same result for $k+l$ odd. If $k<l$ we can simply switch 
$k$ and $l$. For $k=l$ we can simply set them equal. Thus we obtain the 
propagator
\begin{align}\label{xx}
\wick{\c x_k^\mu \c 
x_l^\nu}=\frac{i\tau}{N}\eta^{\mu\nu}\left\{\begin{array}{ll}
l(k-N)&\text{if }k>l\\
k(l-N)&\text{if }k<l\\
\end{array}\right.	.
\end{align}
In the continuum limit with set $t=\tau k$, $t'=\tau l$ and $\tau N=T$, which 
yields
\begin{align}\label{}
\wick{\c x^\mu(t) \c x^\nu(t')}=i\eta^{\mu\nu}\left\{\begin{array}{ll}
\frac{t(t'-T)}{T}&\text{if }t<t'\\
\frac{t'(t-T)}{T}&\text{if }t>t'\\
\end{array}\right. \overset{T\rightarrow 
\infty}{\rightarrow}-i\eta^{\mu\nu}\min(t,t')~,
\end{align}
in agreement with \cref{corr1} in the main text.

All other two-point correlators can be calculated from \cref{xx}.
Let us start with 
\begin{align}
\wick{\c \Delta x^\mu_k\c x^\nu_l}&=\frac{1}{\tau}
\left(\wick{\c x_{k+1}^\mu \c 
x_l^\nu}-\wick{\c x_k^\mu \c x_l^\nu}\right) 
=\frac{i}{ 
N}\eta^{\mu\nu}\left\{\begin{array}{ll}
l\text{ if }k\geq l\\
l-N\text{ if }k<l\\
\end{array}\right.	~,
\label{discr}
\end{align}
which in the continuum limit becomes
\begin{equation}\label{}
\contraction{}{\dot{x}^\mu}{(t)}{x^\nu}\dot{x}^\mu(t) 
x^\nu(t')=i\eta^{\mu\nu}\left\{\begin{array}{ll}
\frac{(t'-T)}{T}&\text{if }t<t'\\
\frac{t'}{T}&\text{if }t\geq t'\\
\end{array}\right. \overset{T\rightarrow 
\infty}{\rightarrow}-i\eta^{\mu\nu}\theta(t'-t)	~,
\end{equation}
in agreement with \cref{corr2} and \cref{corrdot1} in the main text. Note in 
particular that
 the definition $\theta(0)\equiv0$ follows from the limit $N\to 
 \infty$ in  
 \cref{discr}.
 
Similarly,
\begin{align}\label{}
\wick{\c \Delta ^\mu_k\c\Delta 
x^\nu_l}&=\frac{1}{\tau^2}\contraction{}{(x^\mu_{k+1}-x^\mu_k)}{}
{(x^\mu_{k+1}-x^\mu_k)}(x^\mu_{k+1}-x^\mu_k)(x^\mu_{k+1}-x^\mu_k) 
=\frac{i\eta^{\mu\nu}}{\tau 
N}\left\{\begin{array}{ll}
1-N\text{ if }k=l\\
1\text{ else}\\
\end{array}\right.	~,
\end{align}
which in the continuum limit reads 
\begin{equation}\label{}
\contraction{}{\dot{x}^\mu}{(t)}{\dot{x}^\nu}\dot{x}^\mu(t) 
\dot{x}^\nu(t')=i\eta^{\mu\nu}\left\{\begin{array}{ll}
\frac{1}{T}-\delta(0)&\text{if }t=t'\\
\frac{1}{T}&\text{else}\\
\end{array}\right. \overset{T\rightarrow 
\infty}{\rightarrow}-i\eta^{\mu\nu}\delta(t'-t)~,
\end{equation}
in agreement with \cref{corr3} and \cref{corrdot2} in the main text.

\section{Trigonometric identities}
\label{sec:trig}
For completeness, we report here the derivation of some trigonometric identities
used in \cref{sec:x}
\paragraph{Claim 1}
\begin{align}
	\prod_{k=1}^{N-1}\left(1-\cos\left(\frac{k\pi}{N}\right)\right)=\frac{N}{2^{N-1}}~.
	\label{b1}
\end{align}
\begin{proof}
	We start our proof by writing 
	\begin{align}
	x^{2N}-1=(x^2-1)\sum_{k=0}^{N-1}x^{2k}.
	\end{align}
	We also observe that the polynomial $x^{2N}-1$ has roots at 
	$\exp\left(i2\pi 
	\frac{k}{2N}\right)$ for $k\in\left[1,...,2N-1\right]$, and can hence be 
	written 
	as 
	\begin{align}
	&\nonumber\prod_{k=0}^{2N-1}
	\left(x-\exp\left(i2\pi\frac{k}{2N}\right)\right)\\		
=&\nonumber(x^2-1)\prod_{k=1}^{N-1}
\left(x-\exp\left(i2\pi\frac{k}{2N}\right)\right)
\prod_{m=N+1}^{2N-1}\left(x-\exp\left(i2\pi\frac{m}{2N}\right)\right)\\	
\overset{m=-n+2N}{=}&\nonumber(x^2-1)\prod_{k=1}^{N-1}\left(x-\exp\left(i2\pi\frac{k}{2N}\right)\right)\prod_{n=1}^{N-1}\left(x-\exp\left(-i2\pi\frac{n}{2N}\right)\right)\\
	=&(x^2-1)\prod_{m=1}^{N-1}\left(x^2+1-2\cos\left(\frac{\pi 
	m}{N}\right)\right).
	\end{align}
	We thus conclude that 
	\begin{align}
	\sum_{k=0}^{N-1}x^{2k}=\prod_{k=1}^{N-1}\left(x^2+1-2\cos\left(\frac{\pi 
		k}{N}\right)\right),
	\end{align}
which concludes the proof as the claim is the 
	special case where $x=1$.
\end{proof}
\paragraph{Claim 2}
\begin{align}\label{sumcosines}
	\sum_{m=1}^{N-1}\cos\left(\frac{pm\pi}{N}\right)=\left\{\begin{array}{ll}
	N-1&\text{if }p=0\\
	-1&\text{if }p\text{ even}\\
	0&\text{if }p\text{ odd}\\
	\end{array}\right.~.
\end{align}
\begin{proof}
	This proof follows simply from a case selection. 
	If $p=0$ the sum adds the number 1 $N-1$ times. If $p=2l$ we can write 
	\begin{align}
	&\nonumber\sum_{m=1}^{N-1}\cos\left(\frac{2lm\pi}{N}\right)\\
	=&\nonumber\sum_{m=1}^{N-1} \Re \left(\exp\left(i\frac{2\pi 
	lm}{N}\right)\right)\\
	=&\nonumber\Re\left[\sum_{m=0}^{N-1}\left(\exp \frac{2\pi 
	il}{N}\right)^m-1\right]\\
	=&\Re\left[\frac{1-e^{2\pi il}}{1-\exp\frac{2\pi 
	il}{N}}-1\right]=-1.
	\end{align}
	If $p$ is odd the last line is fairly similar
	\begin{align}
	\Re\left[\frac{1-e^{2\pi il}e^{i\pi}}{1-\exp\frac{\pi 
	i(2l+1)}{N}}-1\right]=\Re\left[\frac{2}{1-\exp\frac{\pi 
	i(2l+1)}{N}}-1\right]=0.
	\end{align}
\end{proof}
\paragraph{Claim 3}
\begin{align}\label{sinquotient}
	\sum_{m=1}^{N-1}\frac{\sin\left(\frac{im\pi}{N}\right)
		\sin\left(\frac{jm\pi}{N}\right)}{1-\cos\left(\frac{m\pi}{N}\right)}
	=-\frac{1}{2}\sum_{k=1}^{i+j-1}\min(2j,2i,k)(1+(-1)^{i+j-k})
	+\frac{(N-1)}{2}\min(2i,2j)~.
\end{align}
\begin{proof}
	We start by writing 
	\begin{align}
	1-\cos\left(\frac{m\pi}{N}\right)=2\sin^2\left(\frac{m\pi}{2N}\right).
	\end{align}
	Next we introduce $y=\exp(\frac{i\pi}{2N})$, which allows us to write
\begin{align}
	\frac{\sin\left(\frac{im\pi}{N}\right)}{\sin\left(\frac{m\pi}{2N}\right)}
	=\frac{y^{2im}-y^{-2im}}{y^m-y^{-m}}.
\end{align}
	Next we note that 
	\begin{align}
	y^{2im}-y^{-2im}=(y^m-y^{-m})\sum_{k=0}^{2i-1}y^{m(2i-1-k)}y^{-mk}.
	\end{align}
	Equipped with this we can rewrite
	\begin{align}
	&\nonumber\frac{\sin\left(\frac{im\pi}{N}\right)\sin\left(\frac{jm\pi}{N}\right)}{1-\cos\left(\frac{m\pi}{N}\right)}=\frac{1}{2}\frac{\sin\left(\frac{im\pi}{N}\right)}{\sin\left(\frac{m\pi}{2N}\right)}\frac{\sin\left(\frac{jm\pi}{N}\right)}{\sin\left(\frac{m\pi}{2N}\right)}\\
	=&\frac{1}{2}\sum_{k=0}^{2i-1}\sum_{l=0}^{2j-1}y^{2m(i+j-k-l-1)}.
	\end{align}
	To rewrite this last double sum we need to preform the following reordering:
	\begin{align}
	&\nonumber\sum_{k=0}^{2i-1}\sum_{l=0}^{2j-1}y^{2m(i+j-k-l-1)}\\
	=&\nonumber\begin{array}{cccccc}
	y^{2m(i+j-1)}&+y^{2m(i+j-2)}&+ ... &+ y^{2m(i-j)}&&\\
	&+y^{2m(i+j-2)}&+ ... &+ y^{2m(i-j)}&+y^{2m(i-j-1)}&\\
	&&&...&&\\
	&&+y^{2m(-i+j)}&+...&&+y^{2m(-i-j+1)}
	\end{array}	\\
	&=\sum_{k=1}^{i+j-1}\left(y^{2m(i+j-k)}+y^{-2m(i+j-k)}\right)\cdot 
	\min(k,2i,2j)+(y^0)\cdot \min(2i,2j)~.
	\label{reordering}
	\end{align}
	The equality in the last line of \cref{reordering} can be justified by the 
	following argument. In 
	the line above we have arranged the $2i\times2j$ grid into a parallelogram 
	by shifting each line one term to the right. Thus we have $2i+2j-1$ columns 
	in total. We now need to figure out how many lines are in the $m^{\rm th}$ 
	column. If $2i<2j$ than we can have at most $2i$ lines per column. 
	Alternatively if $2j<2i$ we can have at most $2j$ lines per column. In 
	addition the number of lines grows when moving in from the left, such that 
	the number of columns is $\min(k,2i,2j)$. By letting the sum go to $i+j-1$ 
	we 
	miss the 
	middle column where the exponent is $0$. There are exactly $\min(2i,2j)$ 
	terms of that form.
	Finally, we can sum over $m$ using \cref{sumcosines}, to get
	\begin{align}
	&\nonumber 
	\sum_{m=1}^{N-1}\frac{\sin\left(\frac{im\pi}{N}\right)\sin\left(\frac{jm\pi}{N}\right)}{1-\cos\left(\frac{m\pi}{N}\right)}\\
	=\nonumber&\frac{1}{2}\sum_{m=1}^{N-1}\left[\sum_{k=1}^{i+j-1}2\cos\left(\frac{m(i+j-k)\pi}{N}\right)\min(k,2i,2j)+\min(2i,2j)\right]\\
	=\nonumber&\frac{1}{2}\sum_{k=1}^{i+j-1}\min(2i,2j,k)\left(2N\delta_{0,i+j-k}-1-(-1)^{i+j-k}\right)+\frac{N-1}{2}\min(2i,2j)\\
	=&-\frac{1}{2}\sum_{k=1}^{i+j-1}\min(2i,2j,k)
	\left(1+(-1)^{i+j-k}\right)+\frac{N-1}{2}\min(2i,2j)~.
	\end{align}
	In the last equality we used the fact, that $k\leq i+j-1$ and thus $i+j-k$ 
	can never be 0. 
\end{proof}

\bibliography{ref.bib}


\end{document}